%% file: Kuzkin_fast_Eng_v11.tex
\documentclass[a4paper,12pt]{article}
\usepackage{amsmath}
\usepackage[pdftex]{graphicx}
\usepackage{amsmath}
\usepackage{amsmath,amsthm}
\usepackage{amsfonts}
\input {def_Kuzkin.tex}
\begin{document}
\baselineskip 16pt
\title{Approach to thermal equilibrium in harmonic crystals with polyatomic lattice}
\author{Vitaly A. Kuzkin\footnote{Peter the Great Saint Petersburg Polytechnical University;
              Institute for Problems in Mechanical Engineering RAS;
              e-mail: kuzkinva@gmail.com}}
\maketitle


\begin{abstract}
We study transient thermal processes in infinite harmonic crystals with complex~(polyatomic) lattice. Initially particles have zero displacements and random velocities such that distribution of temperature is spatially uniform. Initial kinetic and potential energies are different and therefore the system is far from thermal equilibrium.
Time evolution of kinetic temperatures, corresponding to different degrees of freedom of the unit cell, is investigated.
It is shown that the temperatures oscillate in time and tend to generally different equilibrium values. The oscillations are caused by two physical processes: equilibration of kinetic and potential energies and redistribution of temperature among  degrees of freedom of the unit cell. An exact formula describing these oscillations is obtained. At large times, a crystal approaches thermal equilibrium, i.e. a state in which the temperatures are constant in time. A relation between  equilibrium values of the temperatures and initial conditions is derived. This relation is refereed to as the non-equipartition theorem. For illustration, transient thermal processes in a diatomic chain and graphene lattice are considered. Analytical results are supported by numerical solution of lattice dynamics equations.

\end{abstract}

\section{Introduction}
In classical systems at thermal equilibrium, the kinetic energy of thermal motion of atoms is usually equally shared among degrees of freedom. This fact follows from the equipartition theorem~\cite{Hoover_stat_phys, Ziman phonon}. The theorem allows to characterize thermal state of the system by a single scalar parameter, notably the kinetic temperature, proportional to kinetic energy of thermal motion.

Far from thermal equilibrium, kinetic energies, corresponding to different degrees of freedom, can be different~\cite{Cases two temp review, Guo2018, Hoover Holian, Holian Mareschal,  Hoovers, laser review}. Therefore in many works several temperatures are introduced~\cite{Cases two temp review, Guo2018, Indeitsev, Petrov Fortov, laser review}. For example, it is well known that temperatures of a lattice and electrons in solids under laser excitation are different~(see e.g. a review paper~\cite{laser review}). Two temperatures are also observed in molecular dynamics simulations of shock waves. In papers~\cite{Holian Mareschal, Hoover Holian, Hoovers, Uribe shockwaves} it is shown that kinetic temperatures, corresponding to thermal motion of atoms  along and across the shock wave front are different. Different nonequilibrium temperatures of sublattices of methylammonium lead halide are reported in papers~\cite{Chang2016, Guo2018}. In papers~\cite{Dhar altern mass, Jou diff mass}, stationary heat transfer in a harmonic diatomic chain connecting two thermal reservoirs is considered.  It is shown that temperatures of sublattices at the nonequilibrium steady state are different.

In the absence of external excitations, the nonequilibrium system tends to thermal equilibrium. Approach to thermal equilibrium is accompanied by several physical processes. Distribution of velocities tends to Gaussian~\cite{Hemmer, Dudnikova_2003, Klein Prigogine, Lanford Lebowitz, Spohn Lebowitz 1977}. The total energy is redistributed among kinetic and potential forms~\cite{Allenbook, Klein Prigogine, Krivtsov 2014 DAN, Kuzkin_FTT}.
Kinetic energy is redistributed between degrees of freedom~\cite{Kuzkin_FTT}. The energy is redistributed between normal modes~\cite{Prigogine normal modes}. These processes, except for the last one, are present in both harmonic and anharmonic systems~\cite{Allenbook, Dudnikova_2003, Klein Prigogine, Krivtsov 2014 DAN, Kuzkin_FTT, Lanford Lebowitz, Spohn Lebowitz 1977}. In harmonic crystals, energies of normal modes do not equilibrate. However distribution of kinetic temperature in infinite harmonic crystals tends to become spatially and temporary uniform~\cite{Hemmer, Spohn Lebowitz 1977, Kuzkin JPhys}.
Therefore the notion of thermal equilibrium is  widely applied to infinite harmonic crystals~\cite{Boldrighini1983, Dobrushin 1986,  Dudnikova_2003, Dudnikova 2005, HEMMEN, Huerta Robertson 1971, Lanford Lebowitz, Spohn Lebowitz 1977, Titulaer}.

Approach to thermal equilibrium in harmonic crystals is studied in  many works~\cite{Boldrighini1983, Dobrushin 1986,  Dudnikova_2003, Dudnikova 2005, HEMMEN, Hemmer, Huerta Robertson 1971, Klein Prigogine, Krivtsov 2014 DAN, Kuzkin_DAN, Kuzkin_FTT, Kuzkin JPhys, Lanford Lebowitz, Linn Robertson 1984, Mielke, Spohn Lebowitz 1977}. Various aspect of this process are studied, including existence of the equilibrium state~\cite{Lanford Lebowitz}, ergodicity~\cite{Titulaer, HEMMEN}, convergence of the velocity distribution function~\cite{Boldrighini1983, Klein Prigogine, Dudnikova_2003, Dudnikova 2005}, evolution of entropy~\cite{Huerta1969, Huerta Robertson 1971, Sokolov} etc. In the present paper, we focus on the behavior of the main observable, notably kinetic temperature~(temperatures).

Two different approaches for analytical treatment of harmonic crystals are presented in literature. One approach employs an exact solution of equations of motion~\cite{Guzev2018, Hemmer, Klein Prigogine, Huerta Robertson 1971, Linn Robertson 1984}. Given known the exact solution, kinetic temperature is calculated as mathematical expectation of corresponding kinetic energy.  For example, in pioneering work of Klein and Prigogine~\cite{Klein Prigogine}, transition to thermal equilibrium in an infinite  harmonic one-dimensional chain with random initial conditions is investigated. Using exact solution derived by Schr\"{o}dinger~\cite{Shredinger}, it is shown that kinetic and potential energies of the chain oscillate in time and tend to equal equilibrium values~\cite{Klein Prigogine}. Another approach uses covariances\footnote{Covariance of two centered random values is equal to mathematical expectation of their product.} of particle velocities and displacements as main variables. For harmonic crystals, closed system of equations for covariances can be derived in steady~\cite{Dhar altern mass, Politi 2008, Lebowitz1967} and unsteady cases~\cite{Krivtsov 2014 DAN, Kuzkin_DAN, Kuzkin_FTT, Kuzkin JPhys, Gavrilov, Politi2010}. Solution of these equations describes, in particular,  time evolution of kinetic temperature. In papers~\cite{Krivtsov 2014 DAN, Kuzkin_DAN, Kuzkin_FTT, Kuzkin JPhys}, this idea is employed for description of approach to thermal equilibrium in harmonic crystals with~{\it simple}~(monoatomic) lattice\footnote{A lattice is referred to as simple lattice, if it coincides with itself under shift by a vector connecting any two particles.}. In particular, monoatomic one-dimensional chains~\cite{Babenkov2016, Krivtsov 2014 DAN} and two-dimensional lattices~\cite{Kuzkin JPhys, Kuzkin_DAN, Kuzkin_FTT} have been covered.

In the present paper, we study approach towards thermal equilibrium in an infinite harmonic crystal with {\it polyatomic} lattice\footnote{Polyatomic lattice consists of several simple monoatomic sublattices. For example, graphene lattice consists of two triangular sublattices.}. Our main goals are to describe time evolution of kinetic temperatures, corresponding to different degrees of freedom of the unit cell, and to calculate equilibrium values of these temperatures.

The paper is organized as follows. In section~\ref{sect EM}, equations of motion for the unit cell are represented in a matrix form. It allows to cover  monoatomic and polyatomic lattices with interaction of an arbitrary number of neighbors and harmonic on-site potential. In section~\ref{sect Transition}, approach to thermal equilibrium is considered. An equation describing the behavior of kinetic temperatures, corresponding to different degrees of freedom of the unit cell, is derived. An exact solution of this equation is obtained. In section~\ref{sect Equilibium}, an expression relating equilibrium values of the temperatures with initial conditions is derived. In sections~\ref{sect Diatomic}, \ref{sect Graphene}, approach to thermal equilibrium in a diatomic chain and graphene lattice are studied. Obtained results are exact in the case of spatially uniform distribution of temperature in harmonic crystals. Implications of the nonuniform temperature distribution and anharmonic effects are discussed in the last section.

\section{Equations of motion and initial conditions}\label{sect EM}
We consider infinite crystals with complex~(polyatomic) lattice in $d$-dimensional space, $d=1,2,3$. In this section, equations of motion of the unit cell are written in a matrix form, convenient for analytical derivations.

Unit cells  of the lattice
are identified by position vectors,~$\xx$, of
their centers\footnote{For analytical derivations, position vectors are more convenient than indices,
because number of indices depends on space dimensionality.}.
Each elementary cell has~$N$ degrees of
freedom~$u_i(\xx), i=1,..,N$, corresponding to components  of particle displacements.
The components of displacements form a column:
\be{}
    \u(\xx) = (u_1, u_2,.., u_N)^{\top},
\ee
where $\top$ stands for the transpose sign.

\begin{figure*}[htb]
\begin{center}
\includegraphics*[scale=0.5]{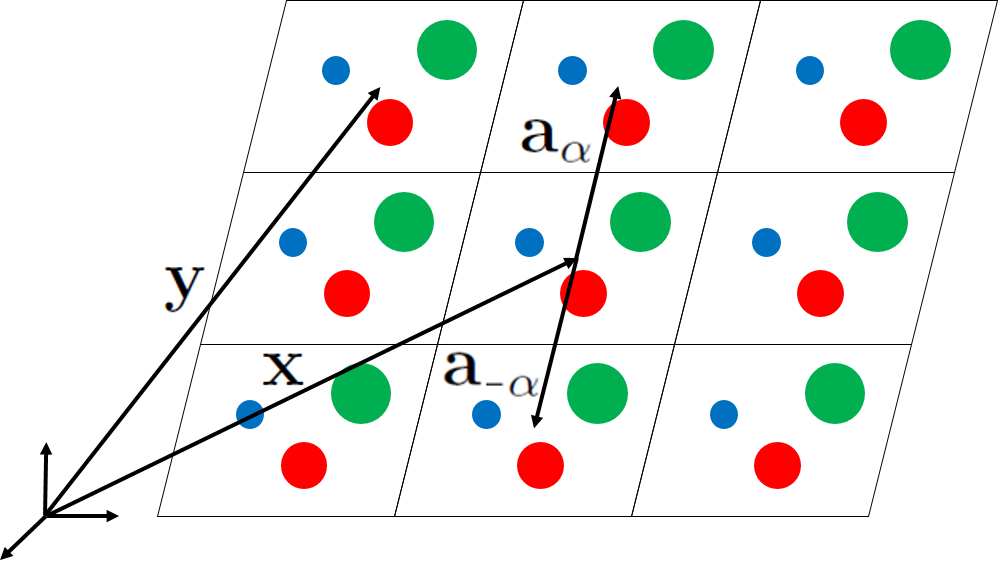}
\caption{Example of a complex two-dimensional lattice with three sublattices.
Particles forming sublattices have different color and size.}
\label{lattice}
\end{center}
\end{figure*}

Particles from the cell~$\xx$ interact with each other and with particles from neighboring unit cells, numbered by index $\alpha$. Vector connecting the cell~$\xx$ with neighboring cell number~$\alpha$ is denoted~$\va$.
Centers of unit cells always form a simple lattice, therefore numbering can be carried out so that vectors $\va$
satisfy the identity:
\be{}
 \va = -\Vect{a}_{-\alpha}.
\ee
Here~$\Vect{a}_{0} = 0$. Vectors $\va$ for a sample lattice are shown in figure~\ref{lattice}.

Consider equations of motion of the unit cell.
In harmonic crystals, the total force acting on each particle is represented as a linear combination of displacements of all other particles.
Using this fact, we write equations of motion in the form\footnote{Similar form of equations of
motion is used in paper~\cite{Mielke}.}:
\be{EM matr}
\begin{array}{l}
\DS \MM \dot{\vv}(\xx) = \sum_{\alpha} \CC_{\alpha} \u(\xx + \va), \qquad \CC_{\alpha} = \CC_{-\alpha}^{\top},
\end{array}
\ee
where~$\vv=\dot{\u}$; $\u(\xx + \va)$ is a column of displacements of particles from unit cell~$\alpha$;  $\MM$ is diagonal $N \times N$ matrix composed of particles' masses; for $\alpha \neq 0$ coefficients of $N \times N$ matrix~$\CC_{\alpha}$ determine stiffnesses of springs connecting unit cell~$\xx$ with neighboring cell number~$\alpha$; matrix~$\CC_{0}$ describes interactions of particles inside\footnote{Additionally, matrix~$\CC_0$ can include stiffnesses of harmonic on-site potential.} the unit cell~$\xx$.  Summation is carried out with respect to all unit cells~$\alpha$, interacting with unit cell~$\xx$~(including~$\alpha=0$).

Formula~\eq{EM matr} describes motion of monoatomic and polyatomic lattices in one-, two-, and three-dimensional cases. For example, one-dimensional diatomic chain and two-dimensional graphene lattice are considered in sections~\ref{sect Diatomic}, \ref{sect Graphene}. Matrices~$\MM$, $\CC_{\alpha}$ for these lattices are given by formulas~\eq{B alp chain}, \eq{Ba graphene}.

{\bf Remark.}   For~$N=1$~(one degree of freedom per unit cell), equation~\eq{EM matr} governs dynamics of the so-called scalar lattices\footnote{In scalar lattices each particle has only one degree of freedom. This model is applicable to monoatomic one-dimensional chains with interactions of arbitrary number of neighbors and to out-of-plane motions of monoatomic two-dimensional lattices.}, considered, for example, in papers~\cite{Harris, Kuzkin JPhys, Mielke, Mishuris}.


The following initial conditions, typical for molecular dynamics modeling~\cite{Allenbook}, are considered:
\be{IC matrix0}
 \u(\xx) = 0, \quad \vv(\xx) = \vv_0(\xx),
\ee
where  $\vv_0(\xx)$ is a column of random initial velocities of particles from unit cell~$\xx$. Components of~$\vv_0(\xx)$ are random numbers with zero mean\footnote{In this case mathematical expectations of all velocities are equal to zero at any moment in time.} and generally different variances. The variances are independent of~$\xx$. Initial velocities of particles in different unit cells are statistically independent, i.e. their covariance is equal to zero. Under these initial conditions, spatial distribution of statistical characteristics, e.g. kinetic temperature, is {\it uniform}.

Equations of motion~\eq{EM matr} with initial conditions~\eq{IC matrix0} completely determine  dynamics of a crystal at any moment in time. The equations can be solved analytically using, for example, discrete Fourier transform. Resulting random velocities of the particles can be used for calculation of statistical characteristics, such as kinetic temperature. However in the following sections, we use another approach, which allows to formulate and solve equations for statistical characteristics of the crystal with {\it deterministic} initial conditions.

\section{Approach to thermal equilibrium}\label{sect Transition}
Initial conditions~\eq{IC matrix0} are such that initially kinetic and potential energies of the crystal are different~(potential energy is equal to zero). Motion of particles leads to redistribution of energy among kinetic and potential forms. Therefore kinetic temperature, proportional to kinetic energy, changes in time. In this section, we derive a formula, exactly describing time evolution of kinetic temperatures, corresponding to different degrees of freedom of the unit cell. The formula shows that a crystal evolves towards a state in which the temperatures are constant in time. This state is further refereed to as the {\it thermal equilibrium}.

\subsection{Generalized kinetic energies. Kinetic temperature}
In this section, we derive an equation, exactly describing time evolution of the kinetic temperatures during approach to thermal equilibrium.

We consider an infinite set of realizations of the same system. The realizations differ only by random initial conditions~\eq{IC matrix0}. This approach allows to introduce statistical characteristics such as kinetic temperatures.

In general, each degree of freedom of the unit cell has its own kinetic
energy and kinetic temperature. Then in order to characterize thermal state of the unit cell, we
introduce~$N\times N$ matrix, $\TT$, further referred
to as the {\it temperature matrix}:
\be{matr temp}
 k_B\TT(\xx) = \MM^{\frac{1}{2}}\av{\vv(\xx) \vv(\xx)^{\top}}\MM^{\frac{1}{2}}
 \quad
 \Leftrightarrow
 \quad
  k_B T_{ij} = \sqrt{M_iM_j} \av{v_i v_j},
\ee
where~$\MM^{\frac{1}{2}}\MM^{\frac{1}{2}} = \MM$; $M_i$ is $i$-th element of matrix~$\MM$, equal to a mass corresponding to i-th degree of freedom;
$k_B$ is the Boltzmann constant; brackets~$\av{..}$ stand for mathematical expectation\footnote{In computer simulations, mathematical expectation can be approximated by average over realizations with different random initial conditions.}.
Diagonal element, $T_{ii}$, of the temperature matrix is equal to kinetic temperature, corresponding to $i$-th degrees of freedom of the unit cell, i.e. $k_B T_{ii} = M_i \av{v_i^2}$. Off-diagonal elements characterize correlation between velocities, corresponding to different degrees of freedom.

We also introduce the kinetic temperature, $T$, proportional
to the total kinetic energy of the unit cell:
\be{kin temp}
  T = \frac{1}{N}\,\tr\TT = \frac{1}{N}\,\sum_{i=1}^N T_{ii},
\ee
where~$N$ is a number of degrees of freedom per unit cell; $\tr(..)$ stands for trace\footnote{Trace of a square matrix is defined as a sum of diagonal elements.} of a matrix. In the case of energy equipartition,  kinetic temperatures, corresponding to all degrees of freedom of the unit cell, are equal to~$T$.

Neither kinetic temperature~\eq{kin temp} nor temperature matrix~\eq{matr temp} is sufficient for derivation of closed system of equations. Therefore we introduce generalized kinetic energy~$\KK(\xx,\yy)$ defined for any pair
of unit cells~$\xx$ and $\yy$ as
\be{cov}
 \KK(\xx, \yy) = \frac{1}{2}\MM^{\frac{1}{2}}\av{\vv(\xx) \vv(\yy)^{\top}}\MM^{\frac{1}{2}}.
\ee
Diagonal elements of matrix~$\KK(\xx,\xx)$ are equal to mathematical expectations of kinetic energies, corresponding to different degrees of freedom of the unit cell, i.e. $K_{ii}(\xx, \xx) = \frac{1}{2}M_i\av{v_i(\xx)^2}$.
The generalized kinetic energy is related to the temperature matrix~\eq{matr temp} as
\be{T vs K}
\frac{1}{2} k_B \TT = \KK(\xx, \xx).
\ee

{\bf Remark.} Notion of generalized kinetic energy for monoatomic lattices is introduced in papers~\cite{Krivtsov 2014 DAN, Kuzkin_DAN, Kuzkin_FTT, Kuzkin JPhys}. In these papers, the energy is defined for pairs of particles rather than pairs of unit cells.

We consider initial conditions~\eq{IC matrix0} such that spatial distribution of all statistical characteristic is uniform. In this case
the following identity is satisfied:
\be{}
    \KK(\xx, \yy) = \KK\(\xx-\yy\).
\ee
Argument~$\xx-\yy$ is omitted below for brevity.
In Appendix I, it is shown that the generalized kinetic energy,~$\KK\(\xx-\yy\)$,
 satisfies equation:
\be{Dyn kappa unif}
\begin{array}{l}
\DS
\ddddot{\KK} -2\(\Dk\ddot{\KK}+\ddot{\KK}\Dk\) + \Dk^2 \KK - 2\Dk\KK \Dk + \KK \Dk^2 = 0,
\\[4mm]
\DS
\Dk \KK \dfeq \suma \MM^{-\frac{1}{2}}\CC_{\alpha}\MM^{-\frac{1}{2}} \KK\(\xx-\yy + \va\).
\end{array}
\ee
Here~$\Dk^2 \KK = \Dk\(\Dk\KK\)$. Formula~\eq{Dyn kappa unif} is equivalent to an infinite system of ordinary differential equations. It {\it exactly} describes the evolution of generalized kinetic energy in any harmonic lattice.

{\bf Remark.} A particular case of equation~\eq{Dyn kappa unif} for a one-dimensional monoatomic harmonic chain with nearest-neighbor interactions was originally derived in paper~\cite{Krivtsov APM 2007}.

{\bf Remark.}  Consider physical meaning of difference operator~$\Dk$. Using this operator, equation of motion~\eq{EM matr} is represented as
\be{}
 \DS \MM^{\frac{1}{2}}  \ddot{\u}(\xx) = \MM^{-\frac{1}{2}}\suma \CC_{\alpha}\u\(\xx + \va\) = \Dk \(\MM^{\frac{1}{2}} \u(\xx)\).
\ee
Therefore~$\Dk$ is equal to operator in the right-hand side  of equations of motion provided that the equations are written for~$\MM^{\frac{1}{2}} \u(\xx)$.

Initial conditions for~$\KK$, corresponding to initial conditions for particle velocities~\eq{IC matrix0}, have form:
\be{IC kapb}
\begin{array}{l}
 \DS \KK = \frac{1}{2} k_B \TT_0 \delta_D(\xx-\yy),
 \qquad
 \dot{\KK} = 0,
 \qquad \ddot{\KK} = \Dk\KK + \KK\Dk,
 \quad  \dddot{\KK} = 0,
\end{array}
\ee
where $\delta_D(0) = 1$; $\delta_D(\xx-\yy)=0$ for~$\xx\neq\yy$; $\TT_0$ is the initial value of temperature matrix~\eq{matr temp}, independent of $\xx$. Here the expression for~$\KK$ follows from formula~\eq{T vs K} and independence of initial velocities of different unit cells. Values~$\dot{\KK}$ and $\dddot{\KK}$ are proportional to covariance of displacements and velocities. Since initial displacements are equal to zero, then initial values of~$\dot{\KK}$, $\dddot{\KK}$ vanish. The expression for~$\ddot{\KK}$ follows from formula~\eq{Dx Dy}, derived in Appendix I.

Thus the generalized kinetic energy~$\KK(\xx-\yy)$ satisfies equation~\eq{Dyn kappa unif} with {\it deterministic} initial conditions~\eq{IC kapb}. Given known the solution of this initial value problem,  the temperature matrix,~$\TT$, is calculated using formula~\eq{matr temp}.

\subsection{Time evolution of the temperature matrix}
In this section, we  solve equation~\eq{Dyn kappa unif} for the generalized kinetic energy with initial conditions~\eq{IC kapb}. The solution yields {\it an exact} expression for the temperature matrix at any moment in time.

The solution is obtained using the discrete Fourier transform with respect to variable~$\xx-\yy$. Vectors~$\xx-\yy$ form the same lattice as
 vectors~$\xx$. Then the vectors are represented as
\be{}
   \xx-\yy = \sum_{j=1}^d a_j z_j \ve_j,
\ee
where $a_j \ve_j, j=1,..,d$ are basis vectors of the lattice; $|\ve_j|=1$; $z_j$ are integers; $d$ is space dimensionality. Direct and inverse discrete Fourier transforms for
an infinite lattice are defined as
\be{DFT}
\begin{array}{l}
\DS \KKh(\kk)
  = \sum_{j=1}^d \sum_{z_j=-\infty}^{+\infty} \KK(\xx-\yy) e^{-\ii \kk \cdot (\xx-\yy)},
 \qquad
   \kk = \sum_{j=1}^d \frac{p_j}{a_j}  \tilde{\ve}_j,
 \\[4mm]
  \DS \KK(\xx-\yy) =  \int_{\kk}  \KKh(\kk) e^{\ii \kk \cdot (\xx-\yy)} {\rm d} \kk.
  \end{array}
\ee
Here $\KKh$ is Fourier image of $\KK$; $\ii^2=-1$; $\kk$ is wave vector;  $\tilde{\ve}_j$ are vectors of the reciprocal
basis, i.e.~$\tilde{\ve}_j\cdot\ve_k = \delta_{jk}$, where $\delta_{jk}$ is the Kroneker delta;
for brevity, the following notation is used:
\be{int k}
\int_{\kk} ... {\rm d}\kk
=
 \frac{1}{(2\pi)^d}\int_{0}^{2\pi}..\int_{0}^{2\pi}  ... {\rm d} p_1.. {\rm d} p_d.
\ee

Applying the discrete Fourier transform~\eq{DFT} in formulas~\eq{Dyn kappa unif}, \eq{IC kapb}, yields
equation
\be{Omega}
\begin{array}{l}
  \DS
\ddddot{\KKh} + 2\(\Omb\ddot{\KKh}+\ddot{\KKh}\Omb\) + \Omb^2 \KKh - 2\Omb\KKh \Omb + \KKh \Omb^2 = 0, \\[4mm]
 \DS \Omb(\kk) =- \sum_{\alpha} \MM^{-\frac{1}{2}}\CC_{\alpha}\MM^{-\frac{1}{2}} e^{\ii\kk\cdot\va},
 \end{array}
\ee
with initial conditions\footnote{Here identities~$\Phi\(\KK(\xx-\yy + \va)\) = \KKh e^{\ii\kk \cdot \va}$, $\Phi(\delta_D(\xx-\yy))=1$, $\Phi\(\Dk\KK\) =-\Omb\KKh$, $\Phi\(\Dk^2\KK\) = -\Omb\Phi\(\Dk\KK\) = \Omb^2\KKh$ are
used.  $\Phi$ is operator of the discrete Fourier transform, i.e. $\Phi\(\KK\)=\KKh$.}
\be{IC kapb image}
\begin{array}{l}
 \DS \KKh = \frac{1}{2} k_B \TT_0, \qquad \dot{\KKh} = 0, \qquad 
 \DS \ddot{\KKh} = -\Omb\KKh - \KKh\Omb, \qquad  \dddot{\KKh} = 0.
\end{array}
\ee
Matrix~$\Omb$ in formula~\eq{Omega} coincides with the  {\it dynamical matrix} of the lattice, derived in Appendix II~(see formula~\eq{Omega app}). Examples of matrix~$\Omb$ for two particular lattices are given by formulas~\eq{Omb chain1}, \eq{Omb graphene}.

To simplify equation~\eq{Omega}, we use the fact that matrix~$\Omb$ is Hermitian, i.e. it is equal to its own conjugate transpose\footnote{Proof of this statement is given in Appendix II.}. Then it can be represented in the form:
\be{PP def}
 \Omb = \PP\Lamb \PP^{*{\top}}, \qquad \Lambda_{ij} = \omega_j^2 \delta_{ij},
\ee
where  $\omega_j^2, j=1,..,N$ are eigenvalues of matrix~$\Omb$ and ~$\omega_j(\kk)$ are branches of dispersion relation for the lattice; $*$ stands for complex conjugate;
matrix\footnote{Matrix $\PP$ is unitary, i.e. $\PP \PP^{*{\top}} = \EE$,  where $\EE$ is identity matrix, i.e. $E_{ij} = \delta_{ij}$}~$\PP$ is composed of normalized eigenvectors of matrix~$\Omb$. Eigenvectors of the dynamical matrix are referred to as polarization vectors~\cite{Dove}. Examples of matrix~$\PP$ are given by formulas~\eq{PP chain}, \eq{PP graphene}.

We substitute formula~\eq{PP def} into~\eq{Omega}.
Then decoupled system of equations with respect to~$\KK'=\PP^{*{\top}}\KKh\PP$ is obtained
\be{kapb'}
\begin{array}{l}
\DS \ddddot{\KK}' + 2\(\Lamb\ddot{\KK}' + \ddot{\KK}'\Lamb\)\
 + \Lamb^2 \KK' - 2\Lamb\KK'\Lamb + \KK' \Lamb^2 = 0
 \Leftrightarrow
 \\[4mm]
 \DS
\Leftrightarrow
 \ddddot{K'}_{ij}
 +
 2 (\omega_i^2 + \omega_j^2)\ddot{K'}_{ij}
 +
 (\omega_i^2 - \omega_j^2)^2 K'_{ij} = 0.
 \end{array}
\ee
Initial conditions for~$\KK'$ are derived by multiplying formulas~\eq{IC kapb image}
 by~$\PP^{*{\top}}$ from the left and by~$\PP$ from the right. Solving equations~\eq{kapb'} with corresponding initial conditions
 and using the relation~\eq{T vs K} for matrices~$\TT$ and $\KK$, yields:
\be{kapb'_{ij}}
\TT = \int_{\kk}\PP \TT'\PP^{*{\top}} {\rm d}\kk,
\qquad
 T'_{ij}
 =
 \frac{1}{2} \{\PP^{*{\top}} \
 \TT_0\PP\}_{ij}
 \[\cos((\omega_i-\omega_j) t)
 +
 \cos((\omega_i+\omega_j) t)\].
\ee
Here $\{...\}_{ij}$ is element~$i,j$ of the matrix. Here and below integration is carried out with respect to dimensionless components of the wave vector~(see formula~\eq{int k}). Formula~\eq{kapb'_{ij}} yields an exact expression for the temperature matrix at any moment in time. 

If initial kinetic energy is equally distributed between degrees of freedom of the unit cell, then~$\TT_0 = T_0 \EE$ and
formula~\eq{kapb'_{ij}} reduces to
\be{TT equip}
 \TT = \frac{T_0}{2}\(\EE  + \int_{\kk} \PP
 \BB(t) \PP^{*\top}{\rm d}\kk\),
\qquad
 B_{ij}(t) = \cos(2\omega_j t) \delta_{ij},
\ee
where~$\EE$ is the identity matrix, i.e. $E_{ij} = \delta_{ij}$.
Formula~\eq{TT equip} shows that during approach to thermal equilibrium temperature matrix is generally not isotropic\footnote{Matrix is called isotropic if it is diagonal and all elements on the diagonal are equal.}, i.e. temperatures, corresponding to  degrees of freedom of the unit cell, are generally different even if their initial values are equal.

Thus formula~\eq{kapb'_{ij}} {\it exactly} describes time  evolution of the  temperature matrix. The majority of further results follow from formula~\eq{kapb'_{ij}}.


\subsection{Time evolution of kinetic temperature}
In this section, using formula~\eq{kapb'_{ij}} we describe the evolution of the kinetic temperature,~$T$, defined by formula~\eq{kin temp}. According to formulas~\eq{matr temp}, \eq{kin temp}, the kinetic temperature is proportional to the total kinetic energy of the unit cell. Since the total energy per unit cell is conserved, then the evolution of the kinetic temperature is caused by redistribution of energy among kinetic and potential forms.

Kinetic temperature is calculated using formula~\eq{kapb'_{ij}}:\footnote{Here the identity~$\tr\(\PP \TT'\PP^{*{\top}}\) = \tr \TT'$ was used.}
\be{T gen}
\begin{array}{l}
  \DS T =
   \frac{T_0}{2}
  \[1
   +
  \frac{1}{N}\sum_{j=1}^{N} \int_{\kk} \(1 + \frac{\{\PP^{*{\top}} \dev\TT_0\PP\}_{jj}}{T_0}\)
  \cos\(2\omega_j(\kk) t\){\rm d} \Vect{k}\],
\end{array}
\ee
where~$\dev\TT_0 = \TT_0 - T_0 \EE$, $T_0=\frac{1}{N} \tr \TT_0$, $\EE$ is identity matrix. Formula~\eq{T gen} shows that evolution of kinetic temperature is influenced by initial distribution of kinetic energy among degrees of freedom of the unit cell. Corresponding example is given in figure~\ref{chain_T_nonequip}.

If initial kinetic energy is equally distributed among degrees of freedom
of the unit cell then~$\dev\TT_0=0$ and formula~\eq{T gen} reduces to
\be{T gen equip}
  T =
  \frac{T_0}{2}
  \[
   1
   +
  \frac{1}{N}\sum_{j=1}^{N} \int_{\Vect{k}} \cos\(2\omega_j(\kk) t\){\rm d} \Vect{k}
  \].
\ee
This expression can also be derived by calculating trace of both parts in formula~\eq{TT equip}.

{\bf Remark.} Formula~\eq{T gen equip} is valid for both monoatomic and polyatomic lattices. It generalizes results obtained in  papers~\cite{Babenkov2016, Krivtsov 2014 DAN,  Kuzkin_DAN, Kuzkin_FTT} for several one-dimensional and two-dimensional monoatomic lattices.
For monoatomic scalar lattices~($N=1$), formula~\eq{T gen equip} reduces to the expression obtained in paper~\cite{Kuzkin JPhys}.

Integrands in formulas~\eq{T gen}, \eq{T gen equip} are rapidly oscillating functions, frequently changing sign inside the integration domain. Integrals of this type usually tend to zero as time tends to infinity~\cite{Fedoryuk}. Therefore kinetic temperature tends to~$\frac{T_0}{2}$. Decrease of temperature is caused by redistribution of energy among kinetic and potential forms. Note that this redistribution is irreversible\footnote{Calculation of change of entropy corresponding to redistribution of energy between kinetic and potential forms would be an interesting extension of the present work.}.

{\bf Remark.} Investigation of asymptotic behavior of integrals~\eq{T gen}, \eq{T gen equip} at large times is not a trivial problem. However, using general results obtained using the stationary phase method~\cite{Fedoryuk}, we can assume that the value~$T-T_0/2$ tends to zero in time as~$1/t^{\frac{d}{2}}$, where $d$ is space dimensionality. Confirmation of this assumption for some particular lattices is given in papers~\cite{Babenkov2016, Krivtsov 2014 DAN, Kuzkin Tsaplin}. Rigorous derivation is beyond the scope of the present paper.

{\bf Remark.} Formulas  \eq{kapb'_{ij}}, \eq{T gen}, \eq{T gen equip} can be generalized for the case of a finite crystal under periodic boundary conditions. In this case, integrals corresponding to inverse discrete Fourier transform are replaced by sums.

Thus time evolution of the temperature matrix, $\TT$, during approach to thermal equilibrium is exactly described by formula~\eq{kapb'_{ij}}. The approach is accompanied by two processes:  oscillations of kinetic temperature, caused by equilibration of kinetic and potential energies~(formulas~\eq{T gen}, \eq{T gen equip}) and  redistribution of kinetic energy between degrees of freedom of the unit cell. From mathematical point of view, the first process is associated with changes of $\tr \TT$, while the second process causes evolution of $\dev\TT$.


\section{Thermal equilibrium. Non-equipartition theorem}\label{sect Equilibium}
\subsection{Random initial velocities and zero displacements}
In this section, we show that the temperature matrix tends to some equilibrium value constant in time. Therefore the notion of thermal equilibrium is used. A formula relating equilibrium value of temperature matrix
with initial conditions is derived using exact solution~\eq{kapb'_{ij}}.

We rewrite formula~\eq{kapb'_{ij}} in the form
\be{TT ij}
\begin{array}{l}
\DS
\TT = \frac{1}{2}
\int_{\kk}\PP
\diag\(\PP^{*{\top}}  \TT_0\PP\)\PP^{*{\top}}{\rm d}\kk
+
\int_{\kk}\PP \tilde{\TT}\PP^{*{\top}} {\rm d}\kk,
\\[4mm]
 \DS \tilde{T}_{ij}
 =
 \frac{1}{2}
 \{\PP^{*{\top}}\TT_0\PP\}_{ij}
 \[\(1-\delta_{ij}\)\cos((\omega_i-\omega_j) t)
 +
 \cos((\omega_i+\omega_j) t)\].
 \end{array}
\ee
Here~$\diag\(...\)$ stands for diagonal part of a matrix. The first term in formula~\eq{TT ij} is independent of time. In the second term, the integrand is a rapidly oscillating function. Such integrals usually asymptotically tend to zero as~$t\rightarrow \infty$~(see e.g. paper~\cite{Fedoryuk}). Therefore the second term vanishes at large times.
Then the temperature matrix tends to equilibrium value,
given by the first term. In order to simplify further analysis, we represent matrix~$\TT_0$ as a sum of isotropic
part and deviator.  Then the first term in formula~\eq{TT ij} reads
\be{HH res2}
\TT_{eq}
   =
   \frac{1}{2N} \tr\(\TT_0\) \EE
   +
   \frac{1}{2}\int_{\kk}\PP \diag\(\PP^{*{\top}} \dev\TT_0 \PP\)\PP^{*{\top}}{\rm d}\kk.
\ee
Formula~\eq{HH res2} relates the equilibrium temperature matrix  with initial conditions. It shows that, in general, equilibrium temperatures, corresponding to different degrees of freedom of the unit cell, are not equal. Formula~\eq{HH res2} is a particular case of the {\it non-equipartition theorem} formulated in the next section.

Formula~\eq{HH res2} shows that if initial kinetic temperatures, corresponding to different degrees of freedom of the unit cell, are equal~($\dev\TT_0=0$) then they are also equal at equilibrium~($\dev\TT_{eq}=0$). Note that during approach to equilibrium the temperature matrix is generally not isotropic, i.e.~$\dev\TT \neq 0$~(see formula~\eq{TT equip}).


\subsection{Arbitrary initial conditions}
In this section, we generalize the results obtained in the previous section for the case of arbitrary initial conditions.

We define generalized potential energy,~$\PI$, generalized Hamiltonian,~$\HH$,
and generalized Lagrangian,~$\LL$ as
\be{HL comp}
\begin{array}{l}
  \DS \HH = \KK + \PI, \qquad \Tens{L} = \KK - \PI,
  \qquad
  \PI= -\frac{1}{4}\(\Dk \DD + \DD \Dk\), \\[4mm]
\DS  \DD(\xx-\yy) = \MM^{\frac{1}{2}} \av{\u(\xx) \u(\yy)^{\top}} \MM^{\frac{1}{2}}.
\end{array}
\ee
In papers~\cite{Krivtsov 2014 DAN, Kuzkin_FTT} similar values are introduced for monoatomic lattices.

Consider relation between generalized kinetic and potential energies at thermal equilibrium. In Appendix I, it is shown that covariance of particle displacements,~$\DD$, and generalized Lagrangian satisfy
the identity
\be{L}
 \LL = \frac{1}{4} \ddot{\DD}.
\ee
We assume that at thermal equilibrium the second time derivative in the right side of formula~\eq{L} is equal to zero. 
Then
equilibrium values of generalized kinetic and potential energies are equal
\be{K Pi H}
 \KK_{eq}= \PI_{eq} = \frac{1}{2}\HH_{eq}.
\ee

Consider a system of equations for equilibrium value of the generalized Hamiltonian,~$\HH_{eq}$. In Appendix III, it is shown that~$\HH$  satisfies additional conservation laws. Writing
the conservation laws for~$\HH_{eq}$, yields
\be{CLaws}
 \tr\HH_{eq} = \tr\HH_0, \qquad \tr\(\Dk^n \dev\HH_{eq}\) = \tr\(\Dk^n \dev\HH_0\),   \qquad   n= 1,2,...
\ee
Here~$\HH_0$ is the initial value of the generalized Hamiltonian.
Also~$\dev\HH$ satisfies equation~\eq{Dyn kappa unif}~(see Appendix~I).
We seek for stationary solution, $\dev\HH_{eq}$, of equation~\eq{Dyn kappa unif}, formulated
for~$\dev\HH$. Then removing time derivatives in this equation yields:  
\be{eq stat}
 \Dk^2 \dev\HH_{eq} - 2\Dk\dev\HH_{eq} \Dk + \dev\HH_{eq} \Dk^2 = 0.
\ee
Solution of equations~\eq{CLaws}, \eq{eq stat} yields equilibrium value of the generalized Hamiltonian,~$\HH_{eq}$. Given known~$\HH_{eq}$, other generalized energies and temperature matrix are calculated using formulas~\eq{T vs K} and~\eq{K Pi H}.

{\bf Remark.} A system of equations, similar to~\eq{CLaws}, \eq{eq stat}, for crystals with monoatomic lattice and interactions of the nearest neighbors was derived in paper~\cite{Kuzkin_FTT}. However solution of this system was obtained only for square and triangular lattices. Here we derive a general solution of system~\eq{CLaws}, \eq{eq stat}, for any polyatomic lattice.

Equations~\eq{CLaws}, \eq{eq stat} are solved as follows. Applying the discrete Fourier transform~\eq{DFT} to these equations
and using formula~\eq{PP def} for matrix~$\Omb$, yields
\be{eq stat 2}
 \Lamb^2 {\HH}' - 2\Lamb {\HH}' \Lamb + {\HH}' \Lamb^2 = 0,
 \qquad \tr\(\Lamb^n {\HH}'\) = \tr\(\Omb^n \dev\hat{\HH}_0\),
 \qquad
 \HH' = \PP^{*{\top}}\dev \hat{\HH}_{eq}\PP.
\ee
Rewriting the first equation from~\eq{eq stat 2} in a component form it can be shown that off-diagonal elements of matrix~${\HH}'$ are equal to zero.
The second equation from~\eq{eq stat 2} is represented as
\be{}
 \sum_{j=1}^{N} \omega_j^{2n} {H}'_{jj} = \sum_{j=1}^{N} \omega_j^{2n} \{\PP^{*{\top}} \dev\hat{\HH}_0 \PP\}_{jj}.
\ee
Then the solution of equations~\eq{eq stat 2} takes form:
\be{H'}
  H'_{ij} = \{\PP^{*{\top}} \dev\hat{\HH}_0 \PP\}_{ij}\delta_{ij}.
\ee
Substitution of formula~\eq{H'} into the last formula from~\eq{eq stat 2} allows to calculate matrix~$\hat{\HH}$. Applying the inverse discrete Fourier transform and using formulas~\eq{T vs K}, \eq{K Pi H},
yields:
\be{HH res}
k_B \TT_{eq} =  \frac{1}{N}\tr\(\HH_0\)\EE
+  \int_{\kk}\PP \diag\(\PP^{*{\top}} \dev\hat{\HH}_0 \PP\)\PP^{*{\top}} {\rm d}\kk.
\ee
In the first term, $\HH_0$ is calculated at~$\xx=\yy$.

{\bf Remark.} Evolution of the generalized Hamiltonian is governed by the forth order equation~\eq{Dyn kappa unif}. Therefore~$\HH_{eq}$ and $\TT_{eq}$, in principle, can be influenced by~$\HH_0$, $\dot{\HH}_0$, $\ddot{\HH}_0$, $\dddot{\HH}_0$. However formula~\eq{HH res} shows that only~$\HH_0$ matters.

Thus formula~\eq{HH res} is a generalization of formula~\eq{HH res2} for the case of arbitrary initial conditions. It shows that, in general, equilibrium temperatures, corresponding to different degrees of freedom of the unit cell, are not equal. Formula~\eq{HH res}  can be referred to as {\it the non-equipartition theorem}. The theorem relates equilibrium temperature matrix with initial conditions.


\section{Example. Diatomic chain}\label{sect Diatomic}
\subsection{Equations of motion}
Presented theory is applicable to crystals with an arbitrary lattice. In this section, the simplest one-dimensional polyatomic lattice is analyzed.

\begin{figure*}[htb]
\begin{center}
\includegraphics*[scale=0.4]{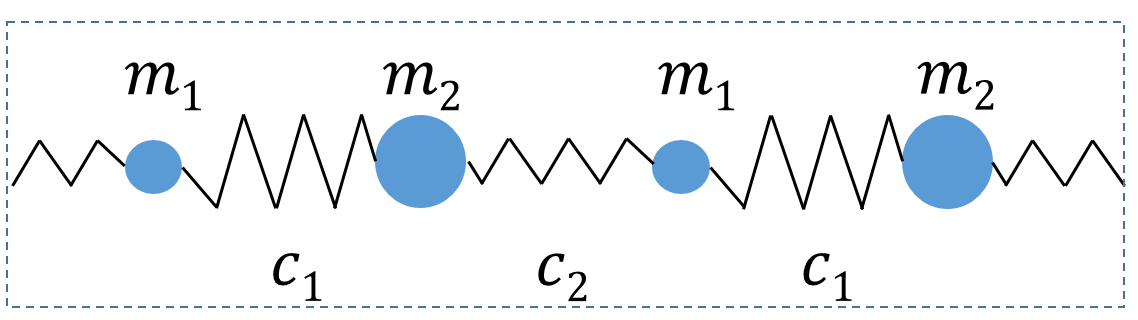}
\caption{Two unit cells of a diatomic chain with alternating masses and stiffnesses. Particles of different size form two sublattices.}
\label{chain_diff_mass_stiff}
\end{center}
\end{figure*}

We consider a diatomic chain with alternating masses~$m_1$, $m_2$ and stiffnesses~$c_1$,
$c_2$~(see fig.~\ref{chain_diff_mass_stiff}). The chain consists of two {\it sublattices}, one formed by particles with masses~$m_1$
and another formed by particles with masses~$m_2$.
This model is frequently used as an example of a system with two branches of  dispersion relation~\cite{Dove, Kosevich, Simon book, Ziman phonon}.

We write equations of  motion of the chain in matrix form~\eq{EM matr}.
Elementary cells, containing two particles each, are numbered by index~$j$. Position vector of the $j$-th unit cell has form:
\be{}
   \xx_j= a j \Vect{e},
\ee
where $a$ is a distance between unit cells, $\ve$ is a unit vector directed along the chain. Each particle has one degree of freedom.
Displacements of particles, belonging to the unit cell~$j$, form a column
\be{}
  \u_j = \u(\xx_j) = (u_{1j},  u_{2j})^{\top},
\ee
where $u_{1j},  u_{2j}$  are displacements of particles with masses~$m_1$ and $m_2$ respectively.
Then equations of motion have form
\be{B alp chain}
\begin{array}{l}
\MM\ddot{\u}_j = \CC_1 \u_{j+1} + \CC_0 \u_{j} + \CC_{-1} \u_{j-1},
\\[4mm]
 \DS \MM = \BM
    m_{1}   &  0 \\
    0       & m_{2}
    \EM,
\quad
\CC_0 =
\BM
-c_1-c_2   &  c_1 \\
c_1 &  -c_1-c_2
\EM,
\quad
\CC_1 = \BM
0   &  0 \\
c_2 &  0
\EM.
\end{array}
\ee
Here~$\CC_{-1} = \CC_{1}^{\top}$.

Initially particles have random velocities and zero displacements.  Velocities of particles with masses~$m_1$, $m_2$ are chosen such that initial temperatures~$T_{11}^0$, $T_{22}^0$ of the sublattices are different. Velocities of different sublattices are uncorrelated, i.e.~$\av{\dot{u}_{1j}\dot{u}_{2j}}=0$.
Then initial temperature matrix has form
\be{IC chain NET}
  \TT_0 = \BM
    T_{11}^0   &  0 \\
    0         &  T_{22}^0
    \EM,
  \qquad
k_B T_{11}^0 = m_1\av{\dot{u}_{1j}^2},
\qquad  k_B T_{22}^0 = m_2\av{\dot{u}_{2j}^2}.
\ee
Here velocities are calculated at $t=0$.
Initial temperature distribution is spatially uniform, i.e.~$T_{11}^0, T_{22}^0$  are independent on~$j$. Further we consider time evolution of temperatures of sublattices, equal to diagonal elements,~$T_{11}$, $T_{22}$, of the temperature matrix.

\subsection{Dispersion relation}
Evolution of temperature matrix is described by formula~\eq{kapb'_{ij}}. In this section, we calculate the dispersion relation and matrix~$\PP$ included in this formula.

\begin{figure*}[htb]
\begin{center}
\includegraphics*[scale=0.35]{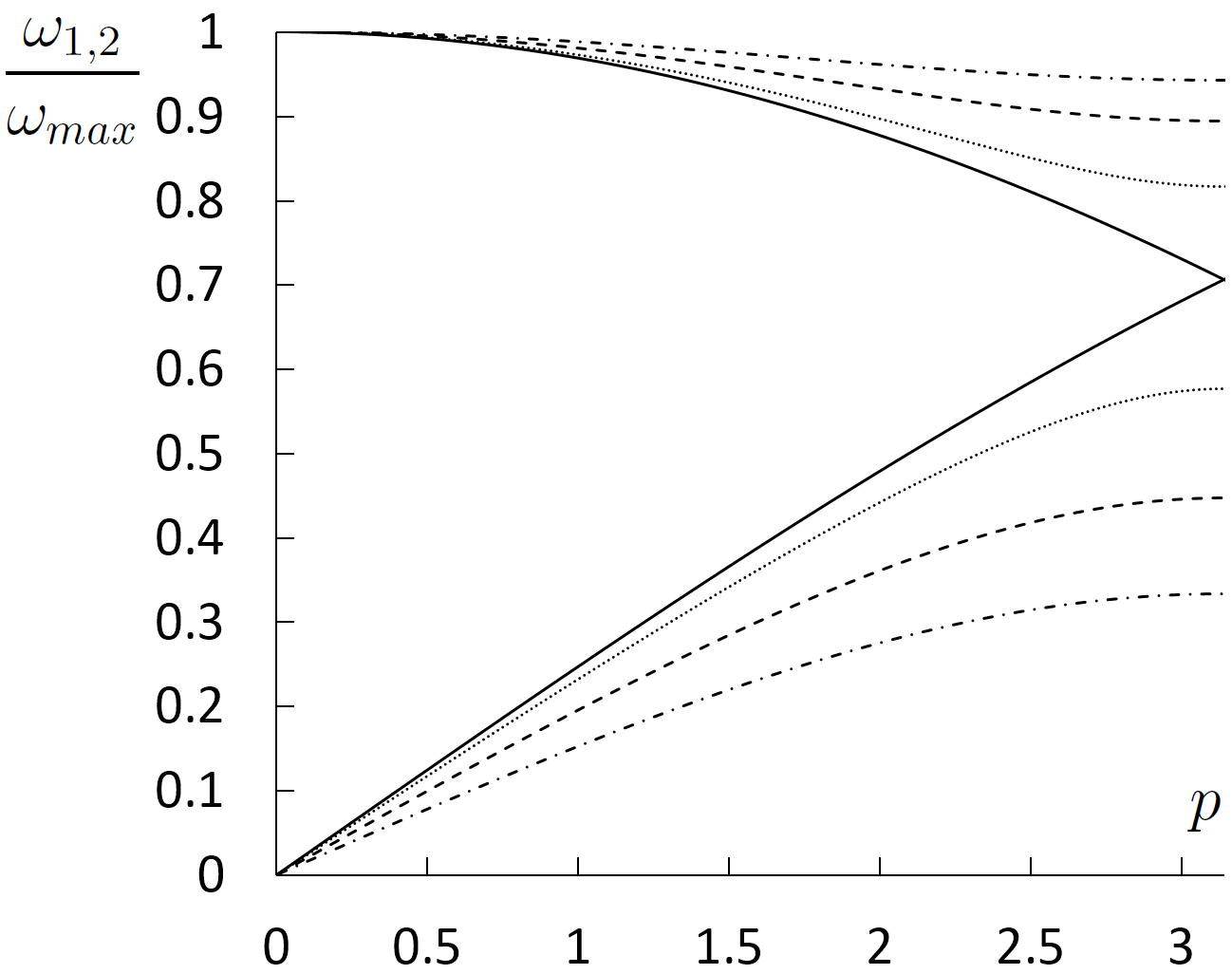}
\caption{Dispersion relation for a chain with alternating stiffnesses~($m_1=m_2$). Curves correspond to different stiffness ratios: $\frac{c_1}{c_2}=1$~(solid line); $\frac{1}{2}$~(dots); $\frac{1}{4}$~(dashed line); $\frac{1}{8}$~(dash-dotted line).}
\label{disp_rel_chain}
\end{center}
\end{figure*}

We calculate  dynamical matrix,~$\Omb$, by formula~\eq{Omega}. Substituting expressions~\eq{B alp chain}
for matrixes~$\CC_{\alpha}$, $\alpha=0; \pm 1$ into formula~\eq{Omega}, we obtain:
\be{Omb chain1}
 \Omb =
 \BM
\frac{c_1+c_2}{m_1}   &  -\frac{c_1 +c_2 e^{-\ii p}}{\sqrt{m_1m_2}} \\
 -\frac{c_1 +c_2 e^{\ii p}}{\sqrt{m_1m_2}}    &  \frac{c_1+c_2}{m_2}
\EM,
\qquad
\kk = \frac{p}{a} \ve,
\ee
where $\kk$ is a wave vector; $p \in [0; 2\pi]$. Calculation of eigenvalues of matrix~$\Omb$,
 yields the dispersion relation:
%
%
%
\be{disp rel chain 2m}
\begin{array}{l}
\DS
\omega^2_{1,2}(p)
=
 \frac{\omega_{max}^2}{2}
  \(1 \pm \sqrt{1 - \frac{16 m_1m_2 c_1c_2 \sin^2\frac{p}{2}}{(m_1+m_2)^2(c_1+c_2)^2}}\),
 \quad
  \omega_{max}^2 = \frac{(c_1+c_2)(m_1+ m_2)}{m_1m_2},
 \end{array}
\ee
where index~$1$ corresponds to plus sign.
Functions~$\omega_{1}(p), \omega_{2}(p)$ are referred to as optical and acoustic branches of the dispersion relation respectively.
Note that~$\omega_{1,2}/\omega_{max}$ equally depend on~$m_1/m_2$ and~$c_1/c_2$. Branches of dispersion relation for different ratios of stiffnesses
are shown in fig.~\ref{disp_rel_chain}.
%

We calculate matrix~$\PP$ in equation~\eq{kapb'_{ij}}. By definition,
matrix~$\PP$ consists of normalized eigenvectors of dynamical matrix~$\Omb$. Eigenvectors~$\Vect{d}_{1,2}$, corresponding to eigenvalues~$\omega_{1,2}^2$~(formula~\eq{disp rel chain 2m}),
have form:
\be{PP chain}
 \Vect{d}_{1,2} = \(1-\frac{m_1}{m_2} \pm \sqrt{\(1-\frac{m_1}{m_2}\)^2 + 4|b|^2\frac{m_1}{m_2}}; -2b\sqrt{\frac{m_1}{m_2}}\)^{\top},
 \qquad b = \frac{c_1 + c_2e^{\ii p}}{c_1 + c_2}.
\ee
Normalization of vectors~$\Vect{d}_{1,2}$ yields columns of matrix~$\PP$.

In the following sections, formulas~\eq{disp rel chain 2m}, \eq{PP chain} are employed for description of temperature oscillations and calculation of equilibrium temperatures of sublattices.

\subsection{Oscillations of kinetic temperature}
In this section, we consider oscillations of kinetic temperature of the unit cell~$T=\frac{1}{2}\(T_{11}+T_{22}\)$. The oscillations are caused by equilibration of kinetic and potential energies.

Initially particles have random velocities and zero displacements. Initial kinetic energies~(temperatures) of sublattices are equal~($T_{11}^0 = T_{22}^0$). The oscillations of kinetic temperature are described by formula~\eq{T gen equip}.
In this case, the formula  reads
\be{chain dif mas temp oscil}
\begin{array}{l}
  \DS   T = \frac{T_0}{2} + T_{ac} + T_{op},
   \quad
    T_{ac} =
    \frac{T_0}{8\pi} \int_0^{2\pi} \cos(2\omega_{2}(p) t){\rm d}p,
    \quad
    T_{op} =
    \frac{T_0}{8\pi} \int_0^{2\pi} \cos(2\omega_{1}(p) t){\rm d}p,
\end{array}
\ee
where $T_0 = \frac{1}{2}\(T_{11}^0 + T_{22}^0\)$ is initial kinetic temperature; dispersion relation~$\omega_j(p), j=1,2$ is given by formula~\eq{disp rel chain 2m}. Contributions of acoustic and optical branches to temperature oscillations are given by integrals~$T_{ac}$, $T_{op}$.

Integrals in formula~\eq{chain dif mas temp oscil} are calculated numerically using Riemann sum approximation. Interval of integration is divided into~$10^3$ equal segments.

\begin{figure*}[htb]
\begin{center}
\includegraphics*[scale=0.33]{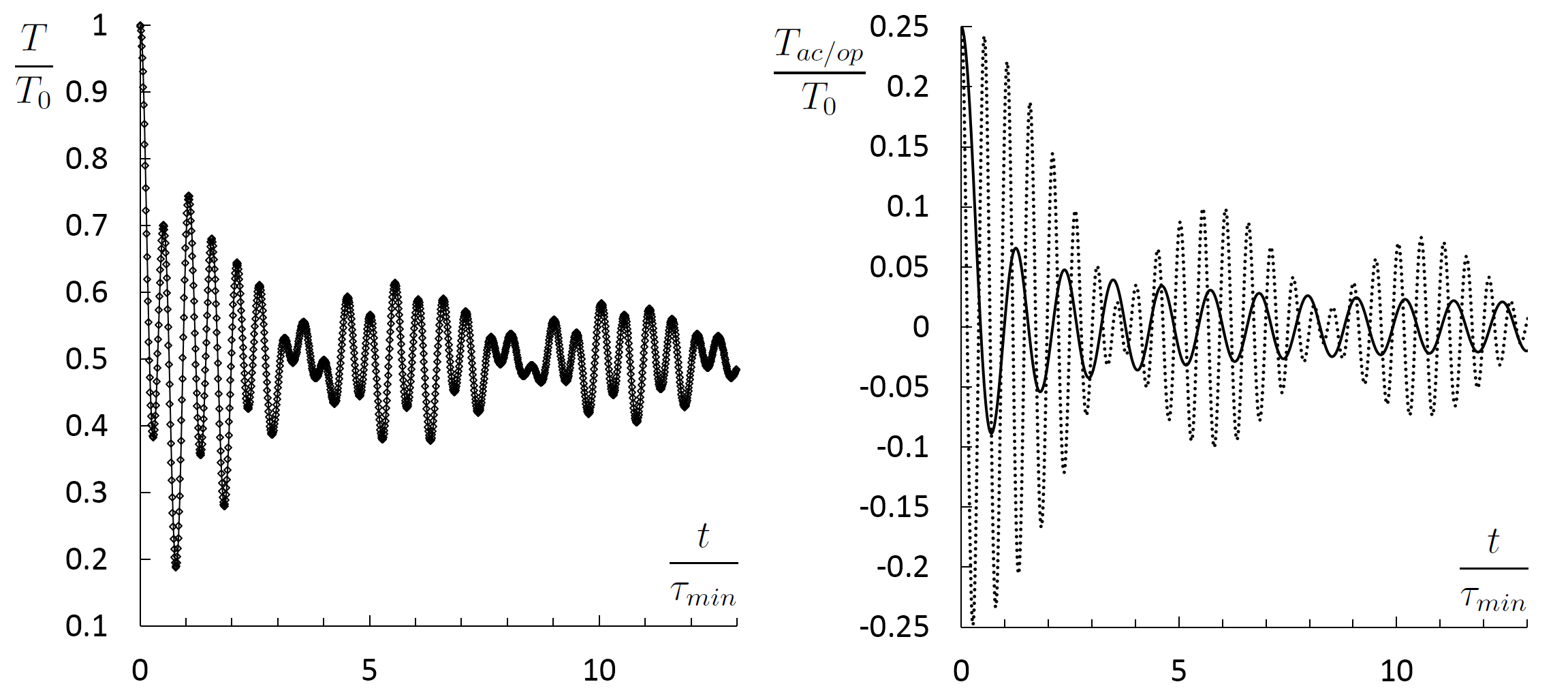}
\caption{A. Oscillations of kinetic temperature~($m_2=4m_1$, $c_1=c_2$). Initial temperatures of sublattices are equal. Analytical solution~\eq{chain dif mas temp oscil}~(solid line), and numerical solution~(dots).
B. Contribution of acoustic~($T_{ac}$, solid line) and optical~($T_{op}$, dots) branches to oscillations of kinetic temperature~($m_2=4m_1$, $c_1=c_2$).}
\label{chain_temp_osc_025}
\end{center}
\end{figure*}

To check formula~\eq{chain dif mas temp oscil}, we compare it with results of numerical solution of lattice dynamics equations~\eq{B alp chain}. In simulations, the chain consists of~$5 \cdot 10^{5}$ particles under periodic boundary conditions.
Numerical integration is carried out using symplectic
 leap-frog integrator with time-step~$10^{-3} \tau_{min}$, where $\tau_{min} = 2\pi/\omega_{max}$, $\omega_{max}$ is defined by formula~\eq{disp rel chain 2m}. During the simulation the total kinetic energy of the chain, proportional to kinetic temperature, is calculated. In this case, averaging over realizations is not necessary.
 Time dependence of temperature for~$m_2/m_1 = 4$ is shown in fig.~\ref{chain_temp_osc_025}A.
%
It is seen that analytical solution~\eq{chain dif mas temp oscil} practically coincides with results of numerical integration of lattice dynamics equations.

Consider contributions~$T_{ac}$, $T_{op}$ of two branches of dispersion relation to oscillations of kinetic temperature.
Time dependencies of~$T_{ac}$, $T_{op}$ for~$m_2=4m_1$,
 $c_1=c_2$ are shown in fig.~\ref{chain_temp_osc_025}B. It is seen that contribution of optical branch has a form of beats~(two close frequencies),
 while contribution of acoustic branch has one main frequency.
 Using the stationary phase method~\cite{Fedoryuk} it can be shown that characteristic frequencies of temperature oscillations belong to frequency spectrum of the chain. Group velocities, corresponding to these frequencies, are equal to zero.
 Figure~\ref{disp_rel_chain} shows that group velocity of acoustic waves is equal to zero for~$p=\pi$, and group velocity of optical waves vanishes at~$p=0, p=\pi$.
 Then main frequencies of temperature oscillations are the following
\be{om123}
\begin{array}{l}
\DS   \omega_{1}|_{p=0}   = \omega_{max},
  \qquad
  \omega_{1}^2|_{p=\pi} = \frac{\omega_{max}^2}{2}
  \(1 + \sqrt{1 - \frac{16 m_1m_2c_1c_2}{(m_1+m_2)^2(c_1+c_2)^2}}\),
  \\[4mm]
  \DS
  \omega_{2}^2|_{p=\pi} = \frac{\omega_{max}^2}{2}
  \(1 - \sqrt{1 - \frac{16 m_1m_2c_1c_2}{(m_1+m_2)^2(c_1+c_2)^2}}\).
\end{array}
\ee
At large times, oscillations of kinetic temperature is represented as a sum of three harmonics with frequencies~\eq{om123} and amplitudes, inversely proportional to~$\sqrt{t}$.

Difference between optical frequencies~$\omega_{1}|_{p=0}$ and $\omega_{1}|_{p=\pi}$ decreases with increasing mass ratio, therefore beats of kinetic
temperature are observed~(see fig.~\ref{chain_temp_osc_025}). Note that similar beats of temperature are observed in two-dimensional triangular lattice~\cite{Kuzkin Tsaplin}.

\begin{figure*}[htb]
\begin{center}
\includegraphics*[scale=0.33]{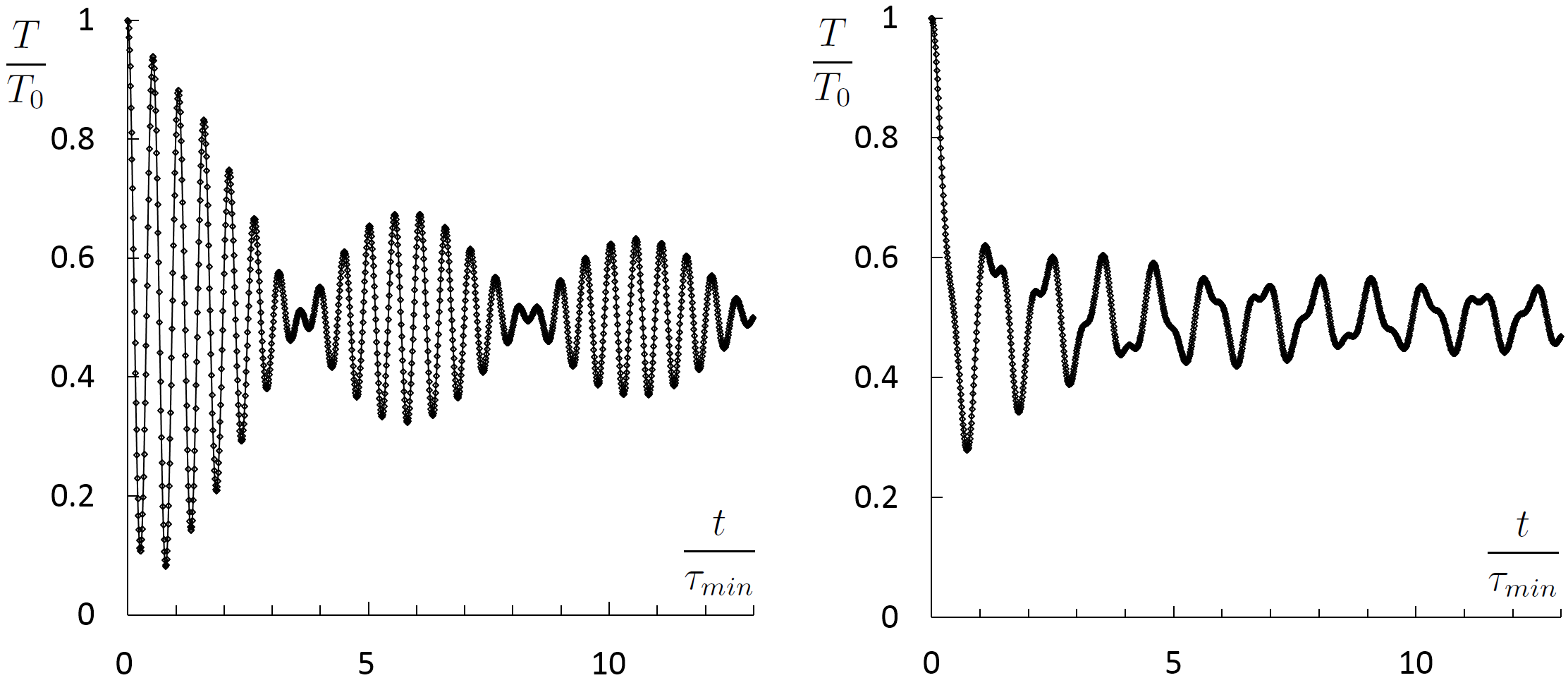}
\caption{Influence of initial temperatures of sublattices on temperature oscillations~($m_2=4m_1$, $c_1=c_2$). Here~$T_{11}^0 \neq 0, T_{22}^0=0$~(left) and $T_{11}^0= 0, T_{22}^0 \neq 0$~(right). Formula~\eq{T gen equip}~(line), and numerical solution of lattice dynamics equations~(dots).}
\label{chain_T_nonequip}
\end{center}
\end{figure*}

Consider influence of ratio between initial temperatures of sublattices~$T_{11}^0$, $T_{22}^0$ on temperature oscillations.
The oscillations for two different cases,~$T_{11}^0 \neq 0, T_{22}^0=0$ and  $T_{11}^0= 0, T_{22}^0 \neq 0$, are shown
in fig.~\ref{chain_T_nonequip}. It is seen that form of oscillations significantly depends on the ratio
between~$T_{11}^0$ and $T_{22}^0$. In both cases, analytical results, obtained using formula~\eq{T gen equip},
practically coincide with numerical solution of lattice dynamics equations.

Thus temperature oscillations are accurately described by formula~\eq{T gen equip}. Amplitude of the oscillations decay in time as\footnote{This fact follows form the asymptotic analysis
based on the stationary phase method~\cite{Fedoryuk}.}~$1/\sqrt{t}$.
 Main frequencies of temperature oscillations belong to spectrum of the chain and correspond to zero group velocities. Form of oscillations significantly depends on initial distribution of energy between sublattices.

\subsection{Redistribution of temperature between sublattices}
%
In this section, we consider the case when initial temperatures of sublattices are not equal~($T_{11}^0 \neq T_{22}^0$). Then temperature is redistributed between the sublattices.

\begin{figure*}[htb]
\begin{center}
\includegraphics*[scale=0.4]{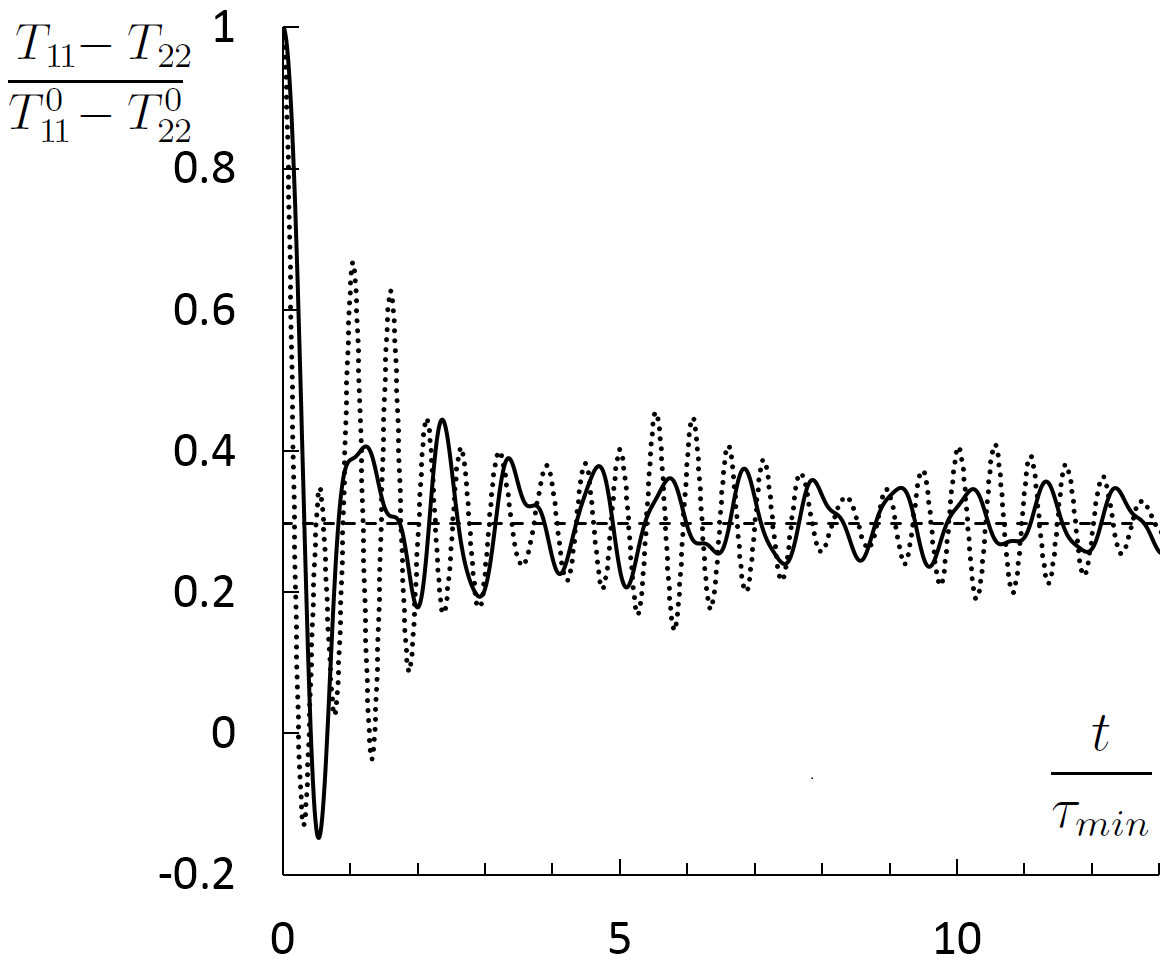}
\caption{Difference between temperatures of sublattices for~$T_{11}^0\neq 0, T_{22}^0=0$~(solid line), and $T_{11}^0= 0, T_{22}^0 \neq 0$~(dots). Here $m_2=4m_1$, $c_1=c_2$, $T_{11}^0, T_{22}^0$ are initial temperatures of sublattices.}
\label{equilibration_diffmasses}
\end{center}
\end{figure*}

Numerical solution of equations of motion~\eq{B alp chain} shows that difference between temperatures of sublattices, $T_{11}-T_{22}$, tends to some equilibrium value. For example, behavior of~$T_{11}-T_{22}$ for~$m_2=4m_1$,
$c_1=c_2$ is shown in  fig.~\ref{equilibration_diffmasses}.
Two cases~$T_{11}^0\neq 0, T_{22}^0=0$ and $T_{11}^0= 0, T_{22}^0 \neq 0$ are considered.
It is seen that in both cases difference between temperatures tends to the value~$0.3(T_{11}^0-T_{22}^0)$,
predicted by formula~\eq{dev KK chain dif mas}. Note that shape of curves for two initial conditions is  different.
Therefore the process of redistribution of temperature between sublattices depends on ratio between~$T_{11}^0$ and $T_{22}^0$.

We calculate the difference between temperatures of sublattices at thermal  equilibrium
using formula~\eq{HH res2}. Deviator of initial temperature matrix has form
\be{IC chain NET2}
  \dev\TT_0 = \frac{T_{11}^0 - T_{22}^0}{2}\Tens{I},
  \qquad
  \Tens{I} =
    \BM
    1   &  0 \\
    0   &  -1
    \EM.
\ee
Substituting~\eq{IC chain NET2} into formula~\eq{HH res2},
yields:
\be{dev KK chain dif mas}
\TT_{eq} = \frac{1}{4}\(T_{11}^0 + T_{22}^0\)\EE
  +
  \frac{T_{11}^0-T_{22}^0}{4\pi}
  \int_0^{2\pi}
   \PP\diag\(\PP^{*{\top}}\Tens{I}\PP\)\PP^{*{\top}} {\rm d} p.
\ee
Matrix~$\PP$ is given by formula~\eq{PP chain}. Formula~\eq{dev KK chain dif mas} yields equilibrium temperatures of sublattices. Integral in formula~\eq{dev KK chain dif mas} is calculated numerically using Riemann sum approximation. Interval of integration is divided into~$10^3$ equal segments.

Consider the case of equal masses~$m_1=m_2$. Using formula~\eq{PP chain} it can be shown that
diagonal elements of matrix~$\PP^{*{\top}}\Tens{I}\PP$ are equal to zero. Then from formula~\eq{dev KK chain dif mas} it follows that for~$m_1=m_2$ and arbitrary~$c_1/c_2$ temperatures of sublattices at thermal equilibrium are equal.

To check formula~\eq{dev KK chain dif mas}, we compare it with results of numerical solution of
lattice dynamics equations~\eq{B alp chain}.
The chain consists of~$10^{4}$ particles under periodic boundary conditions.
We limit ourselves by the following range of parameters: ~$m_1/m_2 \in [0; 1]$
and $c_1/c_2 \in [0;1]$. Numerical integration is carried out with time step~$10^{-3} \tau_*$, where $\tau_*=\sqrt{\frac{c_1+c_2}{m_1}}$. Initially particles have random velocities such that
one of sublattices has zero temperature. During the simulation temperatures of sublattices are calculated.
Equilibrium temperatures  are computed by averaging corresponding kinetic energies over time interval~$[t_{max}/4; t_{max}]$,
where $t_{max}$ is the total simulation time. Reasonable accuracy is achieved for~$t_{max} = 10^2\tau_*$.

\begin{figure*}[htb]
\begin{center}
\includegraphics*[scale=0.32]{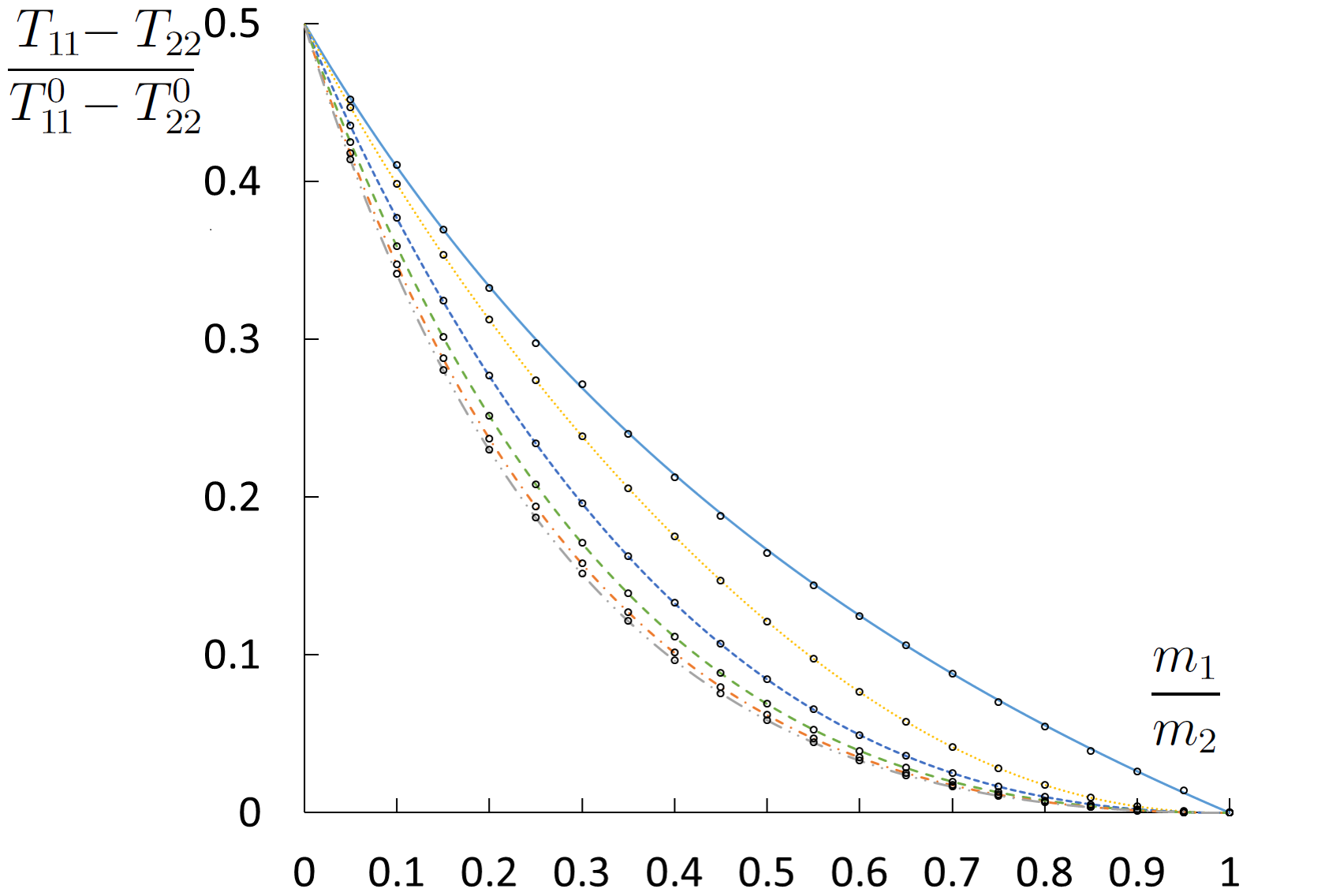}
\caption{Difference between equilibrium temperatures of sublattices for a diatomic chain. Here $T_{11}^0, T_{22}^0$ are initial temperatures of sublattices.
 Curves are calculated using formula~\eq{dev KK chain dif mas}
 for~$\frac{c_1}{c_2}=1$~(solid line); $\frac{1}{2}$~(doted line); $\frac{1}{4}$~(short dashed line); $\frac{1}{8}$~(dashed line); $\frac{1}{16}$~(dash-dotted line); $\frac{1}{32}$~(dash-double doted line).
Circles correspond to results of numerical integration of equations of motion~\eq{B alp chain}.}
\label{stac_diffmasses}
\end{center}
\end{figure*}
%

Equilibrium difference between temperatures of sublattices
for different mass and stiffness ratios is shown in fig.~\ref{stac_diffmasses}.
It is seen that for any given mass ratio,  difference between temperatures decreases with decreasing~$c_1/c_2$ and  tends to a limiting value corresponding to the case~$c_1/c_2 \rightarrow 0$.
In particular, results for~$c_2=64 c_1$ and~$c_2=32 c_1$ are practically indistinguishable.

Thus for the given system,
equilibrium temperatures of sublattices are equal if either
1) initial temperatures are equal~$T_{11}^0=T_{22}^0$ or 2) masses are equal~$m_1=m_2$ and stiffness ratio is arbitrary.
 In general, equilibrium temperatures of sublattices are different. Their values are accurately determined by formula~\eq{dev KK chain dif mas}.

{\bf Remark.} For equal stiffnesses~$c_1=c_2$, equation~\eq{B alp chain} is also valid for
transverse vibrations of a stretched diatomic chain.  In this case, the stiffness is determined by magnitude of
 stretching force. Therefore all results obtained in this section can be used in
 the case of transverse vibrations.

{\bf Remark.} Our results may serve for better understanding of heat transfer in diatomic chains. In papers~\cite{Dhar altern mass, Xiong2013 diff stiff}, stationary heat transfer in diatomic chains connecting two thermal reservoirs with different temperatures was investigated. In paper~\cite{Casher Lebowitz} it was shown that for  $m_1\neq m_2$ temperatures of sublattices were different~(temperature profile was not smooth), while in paper~\cite{Xiong2013 diff stiff} temperatures of sublattices for~$m_1=m_2$ and $c_1\neq c_2$ were practically equal. We suppose  that  these results can be explained using formula~\eq{dev KK chain dif mas}, which  shows that equilibrium temperatures of sublattices are equal only for~$m_1=m_2$.



\section{Example. Graphene lattice~(out-of-plane motions)} \label{sect Graphene}
\subsection{Equations of motion}
\begin{figure*}[htb]
\begin{center}
\includegraphics*[scale=0.25]{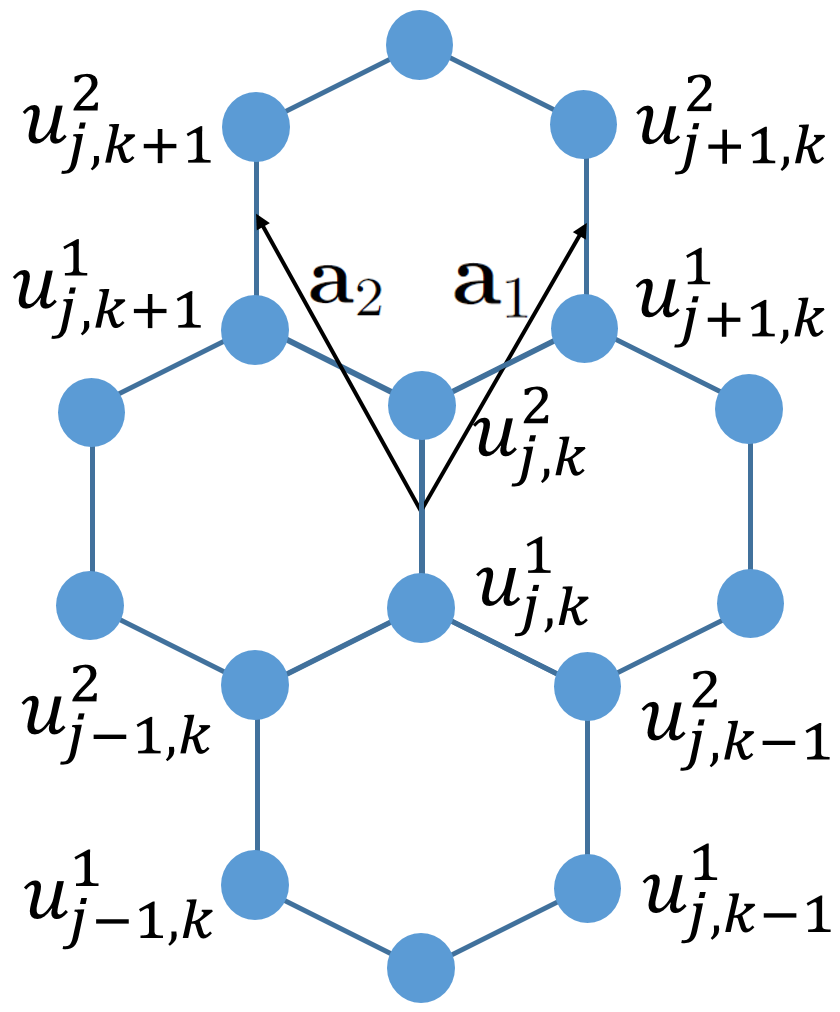}
\caption{Numbering of unit cells and basis vectors~$\Vect{a}_1$, $\Vect{a}_2$ for graphene lattice. Particles move along the normal to lattice plane.}
\label{graphene_lattice}
\end{center}
\end{figure*}

In this section, we consider approach to thermal equilibrium in
hexagonal lattice~(see fig.~\ref{graphene_lattice}). Only out-of-plane vibrations are considered.
The given model describes out-of-plane vibrations of a stretched graphene sheet~\cite{Balandin graphene, Berinskii, Hizhnyakov graphene}.
In-plane vibrations can be considered separately, since in harmonic approximation in-plane and out-of-plane
vibrations are decoupled.

Elementary cells, containing two particles each, are numbered by a pair of indices~$j,k$~(see fig.\ref{graphene_lattice}).
Basis vectors~$\Vect{a}_1$ and~$\Vect{a}_2$ for graphene have form:
\be{}
   \Vect{a}_1 = \frac{\sqrt{3}a}{2} \( \Vect{i} + \sqrt{3}\Vect{j} \),
   \qquad
   \Vect{a}_2 = \frac{\sqrt{3}a}{2} \(\sqrt{3}\Vect{j} - \Vect{i}\),
\ee
where $\Vect{i}, \Vect{j}$ are Cartesian  unit vectors; in fig.~\ref{graphene_lattice}
vector~$\Vect{i}$ is horizontal.
Vector~$\Vect{a}_1$ connects centers of cells~$j,k$ and $j+1,k$.
 Vector~$\Vect{a}_2$ connects centers of cells~$j,k$ and $j,k+1$.
 Position vector of cell~$j,k$ has form:
\be{}
  \xx_{j,k} = \sqrt{3}a \(j \Vect{e}_1 + k  \Vect{e}_2\),
  \qquad
  \Vect{e}_1 = \frac{\Vect{a}_1}{|\Vect{a}_1|},
  \qquad
  \Vect{e}_2 = \frac{\Vect{a}_2}{|\Vect{a}_2|}.
\ee

Each particle has one degree of freedom~(displacement normal to lattice plane).
 Displacements of a unit cell~$j,k$ form a column:
\be{}
 \u_{j,k} = \u(\xx_{j,k}) = \(u^1_{j,k}, u^2_{j,k}\)^{\top},
\ee
where~$u^1_{j,k}, u^2_{j,k}$ are displacements of two sublattices.

Consider equations of motion of unit cell~$j,k$. Each particle is connected with
three nearest neighbors by linear springs~(solid lines in fig.~\ref{graphene_lattice}).
Equilibrium length of the spring is less than initial distance between particles, i.e. the graphene sheet is uniformly
stretched~\footnote{In the absence of stretching, out-of-plane vibrations are essentially nonlinear.
Various nonlinear effects in unstrained graphene are considered e.g. in papers~\cite{Dmitriev1, Dmitriev2}.}. Stiffness of the spring, determined by stretching force, is denoted by~$c$.
Then equations of motion have form
\be{Ba graphene}
\begin{array}{l}
\MM\ddot{\u}_{j,k}
=
\CC_{1} \u_{j+1,k} + \CC_{-1} \u_{j-1,k}
+ \CC_{0} \u_{j,k} + \CC_{2} \u_{j,k+1} + \CC_{-2} \u_{j,k-1},
\\[4mm]
\CC_0 =
\BM
-3c &  c \\
c   &  -3c
\EM,
\quad
\CC_1 = \CC_2 = \BM
0  &  0 \\
c  &  0
\EM,
\quad
\MM = \BM
m  &  0 \\
0  &  m\EM.
\end{array}
\ee
Here~$\CC_{-1} = \CC_{1}^{\top}$, $\CC_{-2} = \CC_{2}^{\top}$; $m$ is particle mass.

Initially particles have random  velocities and zero displacements.
Velocities are chosen such that initial temperatures
 of  sublattices are different~($T_{11}^0 \neq T_{22}^0$). Velocities of different sublattices are uncorrelated, i.e.~$\av{\dot{u}^1_{j,k}\dot{u}^2_{j,k}}=0$.
Then initial temperature matrix,~$\TT_0$, has form:
\be{IC gr}
\begin{array}{l}
\DS \TT_0 = \BM
    T_{11}^0   &  0 \\
    0   &  T_{22}^0
    \EM,
\qquad
    k_B T_{11}^0 = m\av{\(\dot{u}^1_{j,k}\)^2},  
    \qquad  k_B T_{22}^0 = m\av{\(\dot{u}^2_{j,k}\)^2}.
  \end{array}
\ee
Here velocities are calculated at $t=0$. Initial temperature distribution is spatially uniform, i.e.~$T_{11}^0, T_{22}^0$  are independent on~$j,k$.
Further we consider time evolution of temperatures of sublattices, equal to diagonal elements of the temperature matrix~$T_{11}$, $T_{22}$.

\subsection{Dispersion relation}\label{sect disp rel graphene}
Evolution of temperature matrix during approach to thermal equilibrium is described by formula~\eq{kapb'_{ij}}.
In this section, we calculate the dispersion relation and matrix~$\PP$ included in this formula.

We calculate dynamical matrix~$\Omb$ using formula~\eq{Omega}. Substituting expressions~\eq{Ba graphene} for matrixes~$\CC_{\alpha}$,
$\alpha=0;\pm1; \pm2$ into formula~\eq{Omega}, we obtain:
\be{Omb graphene}
\begin{array}{l}
\DS
 \Omb = \omega_*^2
            \BM
            3                   &  -1 -e^{-\ii p_1}-e^{-\ii p_2} \\
            -1 -e^{\ii p_1}-e^{\ii p_2}   &  3
            \EM,
            \qquad
            p_1 = \kk\cdot \Vect{a}_1, \quad p_2 = \kk\cdot \Vect{a}_2,
\end{array}
\ee
where~$\kk$ is wave-vector; $\omega_*^2 = \frac{c}{m}$;
$p_1, p_2 \in [0; 2\pi]$ are dimensionless components of the wave vector.

\begin{figure*}[htb]
\begin{center}
\includegraphics*[scale=0.3]{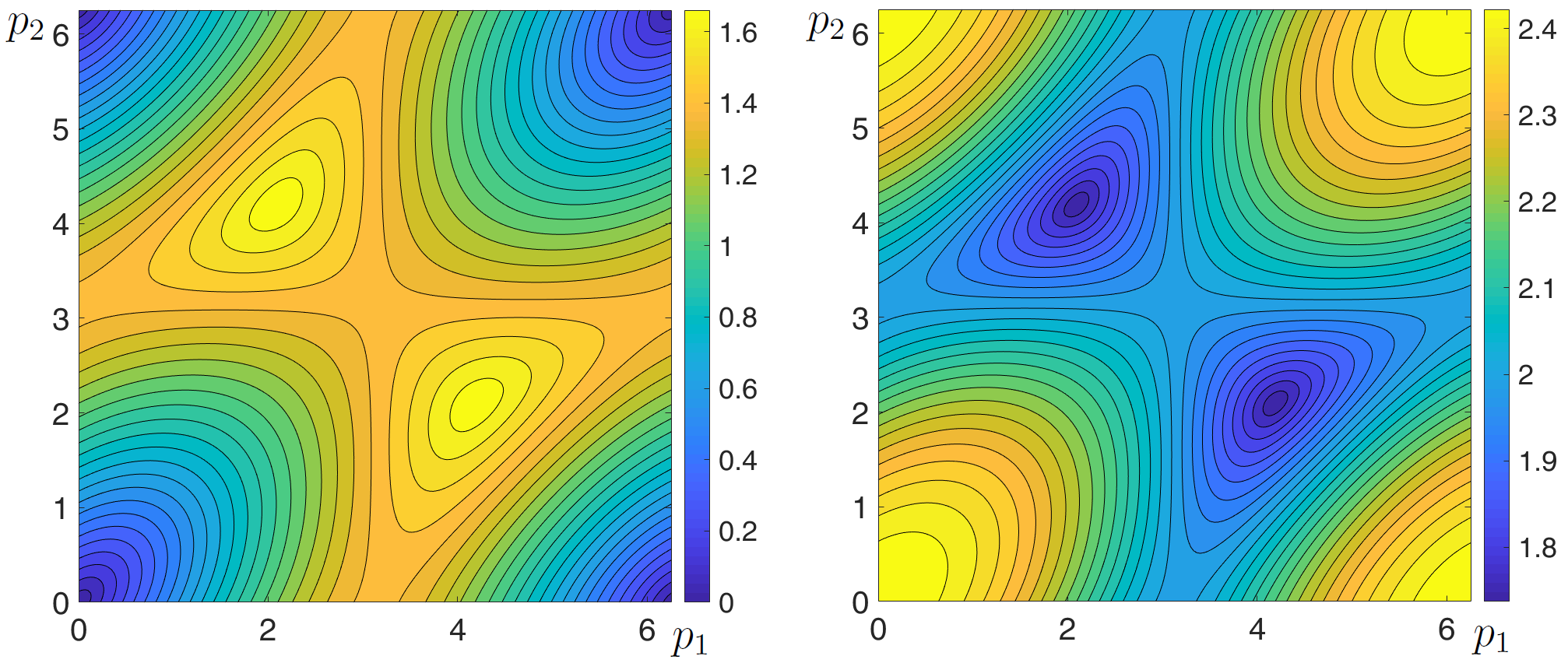}
\caption{Acoustic~($\omega_{2}(p_1,p_2)/\omega_*$, left)
 and optical~($\omega_{1}(p_1,p_2)/\omega_*$, right) dispersion surfaces~\eq{disp gr}
 for out-of-plane vibrations of graphene.}
\label{disp_rel_graphene}
\end{center}
\end{figure*}

Eigenvalues~$\omega_{1}^2, \omega_{2}^2$ of matrix~$\Omb$ determine dispersion relation for the lattice.
Solution of the eigenvalue problem yields:
%
%
\be{disp gr}
    \omega_{1,2}^2 = \omega_*^2
    \(3 \pm \sqrt{3 + 2 \(\cos p_1 + \cos p_2 + \cos\(p_1-p_2\)\)}\),
\ee
where index~$1$ corresponds to plus sign.
Functions~$\omega_{1}(p_1,p_2)$, $\omega_{2}(p_1,p_2)$ are referred to as optical and acoustic dispersion surfaces
respectively~(see fig.~\ref{disp_rel_graphene}).
Eigenvectors of matrix~$\Omb$ are columns of matrix~$\PP$:
\be{PP graphene}
 \PP =
    \frac{1}{\sqrt{|b|^2 + b^2}}
    \BM
    |b|   &  |b| \\
    -b    &   b
    \EM,
 \qquad
 b = 1 + e^{\ii p_1} + e^{\ii p_2}.
\ee

In the following sections, formulas~\eq{Omb graphene}, \eq{disp gr},  \eq{PP graphene} are employed for description of temperature oscillations and calculation of equilibrium temperatures of sublattices.

\subsection{Oscillations of kinetic temperature}
In this section, we consider oscillations of kinetic temperature of the unit cell~$T = \frac{1}{2}(T_{11} + T_{22})$ in graphene.

In general, the oscillations are described by formula~\eq{T gen}. Using formulas~\eq{IC gr}, \eq{PP graphene}
it can be shown that diagonal elements of matrix~$\PP^{*{\top}} \dev\TT_0 \PP$ are equal to zero. Then from formula~\eq{T gen} it follows that temperature oscillations are independent of the ratio between
temperatures of sublattices~$T_{11}^0$ and $T_{22}^0$. Then formula~\eq{T gen equip} can be used:
\be{T graphene}
\begin{array}{l}
  \DS   T = \frac{T_0}{2} + T_{ac} + T_{op},
 \qquad
  T_{ac} = \frac{T_0}{16\pi^2} \int_0^{2\pi} \int_0^{2\pi} \cos(2\omega_{2}(p_1,p_2) t) {\rm d}p_1{\rm d}p_2,
  \\[4mm]
  \DS T_{op} = \frac{T_0}{16\pi^2} \int_0^{2\pi} \int_0^{2\pi} \cos(2\omega_{1}(p_1,p_2) t) {\rm d}p_1 {\rm d}p_2,
\end{array}
\ee
where $T_0 = \frac{1}{2}\(T_{11}^0 + T_{22}^0\)$ is initial kinetic temperature;
functions~$\omega_{1,2}(p_1,p_2)$ are given by formula~\eq{disp gr}.
In further calculations, integrals in formula~\eq{T graphene} are evaluated using Riemann sum approximation.
Integration area is divided into~$400\times 400$ equal square elements.

\begin{figure*}[htb]
\begin{center}
\includegraphics*[scale=0.32]{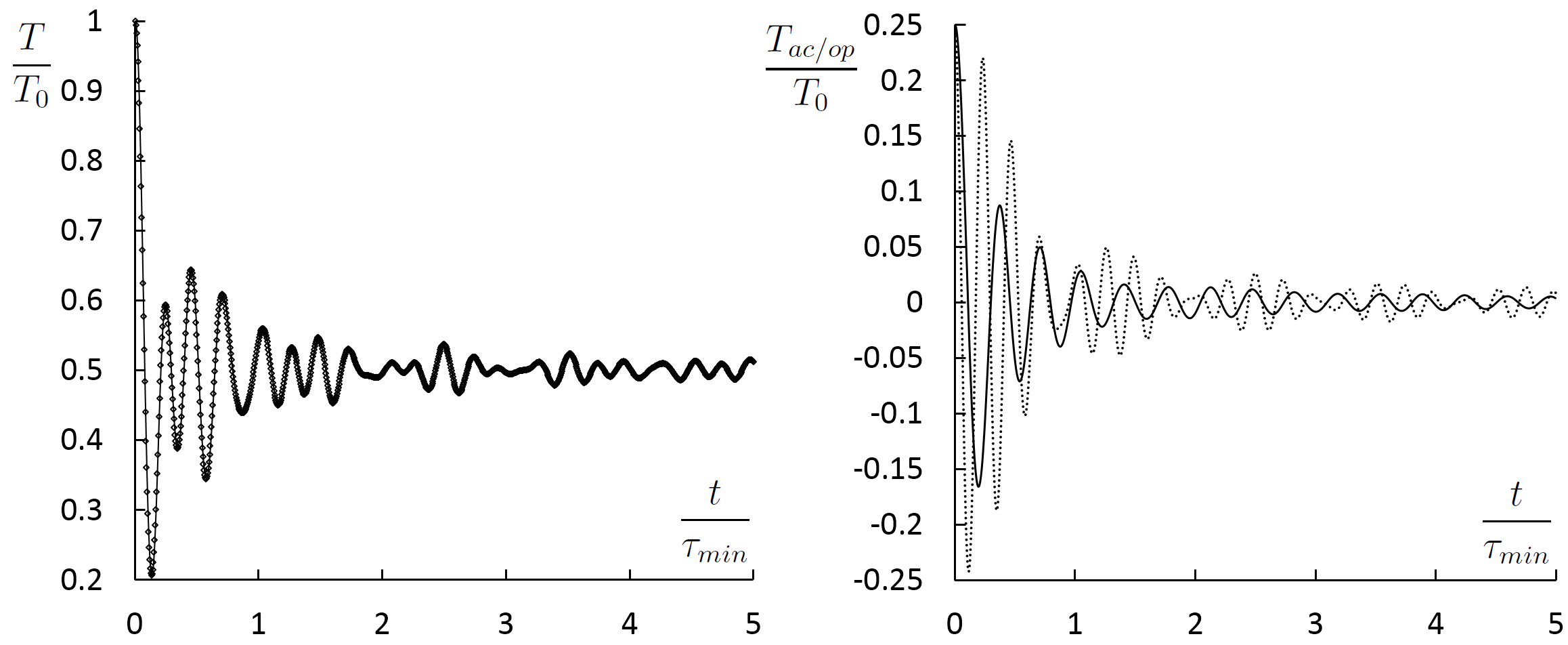}
\caption{A. Oscillations of kinetic temperature in graphene sheet with random initial velocities and zero displacements. Numerical
solution of equations of motion~\eq{Ba graphene}~(dots) and analytical solution~\eq{T graphene}~(line).
B. Contribution of acoustic~($T_{ac}$, solid line) and optical~($T_{op}$, dotted line)
dispersion surfaces to temperature oscillations in graphene.}
\label{graphene_temp_osc}
\end{center}
\end{figure*}

To check formula~\eq{T graphene}, we compare it with results of numerical solution of lattice
dynamics equations~\eq{Ba graphene}. In our simulations, graphene sheet contains~$10^3 \times 10^3$ unit cells under periodic boundary conditions.
Numerical integration is carried out with time-step~$5 \cdot 10^{-3} \tau_{*}$, where $\tau_{*} = 2\pi/\omega_{*}$.
 During the simulation the total kinetic energy of the lattice, proportional to kinetic temperature, is calculated.
 In this case, averaging over realizations is not necessary. Time dependence of temperature
 is presented in fig.~\ref{graphene_temp_osc}A.
The figure shows that formula~\eq{T graphene} accurately describes temperature oscillations.
Calculations with different initial temperatures of sublattices~($T_{11}^0 \neq T_{22}^0$)  confirm out  conclusion that the ratio of these temperatures do not influence the behavior of~$T$.

Contributions of acoustic and optical dispersion surfaces to temperature oscillations are shown in fig.~\ref{graphene_temp_osc}B.
The contributions are given by integrals~$T_{ac}$ and $T_{op}$~(formula~\eq{T graphene}).
It is seen that oscillations corresponding to optical dispersion surface has two main frequencies, while oscillations corresponding to acoustic surface has only one main frequency.
The frequencies can be calculated using asymptotic analysis of integrals~\eq{T graphene} at large~$t$ using the stationary phase
method~\cite{Fedoryuk}. This investigation is beyond the scope of the present paper. Similar investigation for two-dimensional
triangular lattice is carried out in paper~\cite{Kuzkin Tsaplin}.

Thus oscillations of kinetic temperature are accurately described by formula~\eq{T graphene}. Amplitude of these oscillations decays in time as~$1/t$.
Formula~\eq{T graphene} is valid for an arbitrary ratio of initial temperatures of sublattices.

\subsection{Redistribution of temperature between sublattices}
In this section, we consider redistribution of kinetic temperature between sublattices in graphene in the case~$T_{11}^0 \neq T_{22}^0$.

Equilibrium temperatures of sublattices are calculated using formula~\eq{HH res2}.
Corresponding expression for initial temperature matrix is given by
formula~\eq{IC gr}. In the previous section, it is mentioned that for graphene,
 matrix~$\PP^{*{\top}}\dev{\TT}_0\PP$ in formula~\eq{HH res2} has zero diagonal elements.
 Then from formula~\eq{HH res2} it follows that~$\dev\TT_{eq}=0$, i.e. temperatures of sublattices equilibrate.

\begin{figure*}[htb]
\begin{center}
\includegraphics*[scale=0.5]{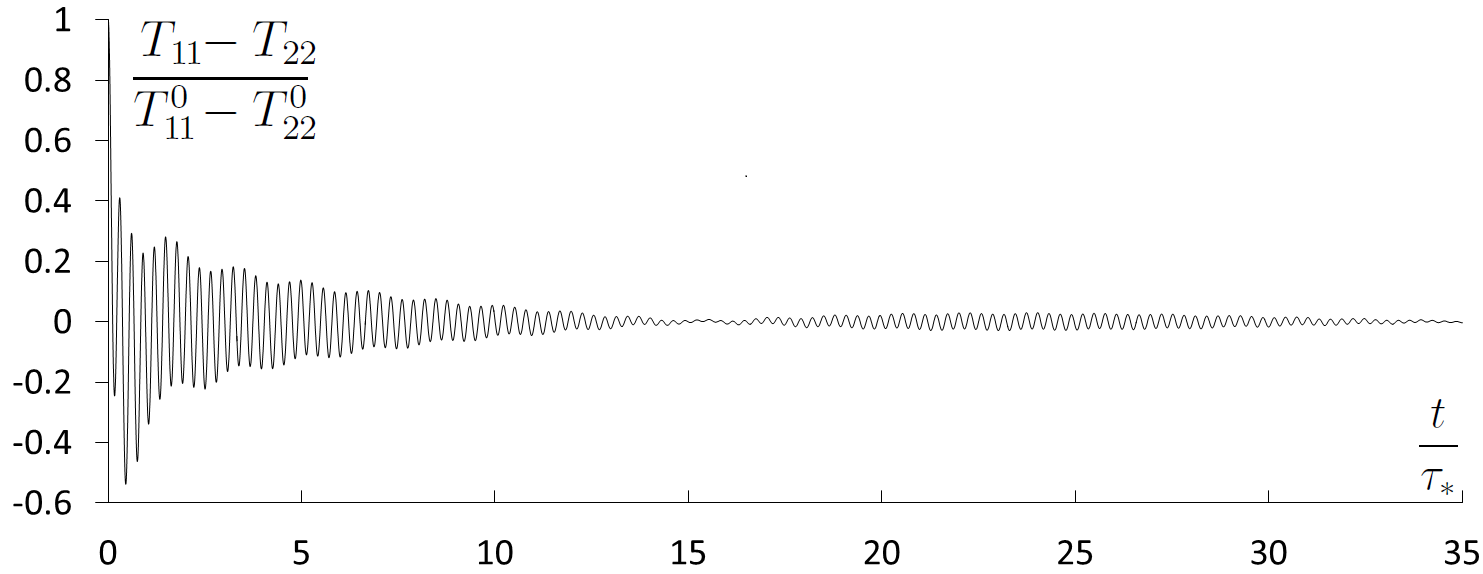}
\caption{Redistribution of kinetic temperatures between sublattices in graphene~(numerical solution of lattice dynamics equations~\eq{Ba graphene}).}
\label{graphene_TxTy}
\end{center}
\end{figure*}

To check this fact, consider numerical solution of equations of motion~\eq{Ba graphene}. Periodic cell containing~$10^3 \times 10^3$
unit cells is used. Initially particles of one sublattice have random velocities, while particles of another sublattice are motionless.
Initial displacements are equal to zero. Numerical integration is carried with time step~$5 \cdot 10^{-3} \tau_*$, where $\tau_*=2\pi/\omega_*$.  Time evolution of temperature difference,~$T_{11}-T_{22}$, is shown
in fig.~\ref{graphene_TxTy}.
%
The figure shows beats of difference between temperatures of sublattices.
The amplitude of beats decays in time as~$1/t$.

Thus at large times, temperatures of sublattices in graphene become equal.

\section{Conclusions}
An analytical description of approach to thermal equilibrium in infinite harmonic crystals with complex~(polyatomic) lattice was presented.

Initially the crystal is in a nonequilibrium state such that
kinetic and potential energies are not equal.
The crystal tends to thermal equilibrium, i.e. to a state in which temperatures, corresponding to different degrees of freedom of the unit cell, are  constant in time. Approach to thermal equilibrium is accompanied by oscillations of the temperatures {\it exactly} described by formula~\eq{kapb'_{ij}}. The oscillations are caused by two physical processes:
1) equilibration of kinetic and potential energies, and
2) redistribution of kinetic energy~(temperature) among degrees of freedom of the unit cell. In $d$-dimensional crystal, amplitude of the oscillations decays in time as~$1/t^{\frac{d}{2}}$.

At large times, kinetic and potential energies equilibrate. Kinetic energy is redistributed between degrees of freedom of the unit cell. Equilibrium values of kinetic temperatures, corresponding to different degrees of freedom of the unit cell, are related with initial conditions by the non-equipartition theorem~(formulas~\eq{HH res2}, \eq{HH res}). The theorem shows that these kinetic temperatures are equal at thermal equilibrium if their initial values are equal. If initial kinetic temperatures are different then they are usually different at equilibrium, except for some lattices. For example, it is shown that in diatomic chain with alternating stiffnesses~(equal masses) and graphene lattice performing out-of-plane motions the equilibrium values of the kinetic temperatures are equal.

Our analytical results are exact in the case of  spatially uniform distribution of kinetic temperatures. In the case of nonuniform temperature distribution, ballistic heat transfer should be considered along with transient
processes described above. However the heat transfer is much slower than the transient processes~\cite{Gavrilov, Krivtsov DAN 2015, Kuzkin JPhys,  Sokolov}. Therefore at small times, the crystal locally almost achieve thermal equilibrium. This almost equilibrium state slowly changes due to ballistic heat transfer. Therefore our results can be used for description of fast local transition to thermal equilibrium in nonuniformly heated crystals.

In the present paper, anharmonic effects were neglected. Anharmonicity leads, in particular, to exchange of energy between normal modes.
In this case, temperatures, corresponding to different degrees of freedom of the unit cell,
tend to equal equilibrium values.
However in papers~\cite{Benettin slow relax, Kuzkin_FTT, Marcelli slow relaxation},
it is shown that, at least in the case of small anharmonicity, the exchange between normal modes is significantly
slower than transient thermal processes described above. Therefore at small times,  transient thermal processes are well described by harmonic approximation.

\section{Acknowledgements}
The author is deeply grateful to A.M. Krivtsov, S.V. Dmitriev, M.A. Guzev, D.A. Indeitsev, E.A. Ivanova, S.N. Gavrilov, I.E. Berinskii and A.S. Murachev for useful discussions. The work was financially supported by the Russian Science Foundation
under grant No. 17-71-10213.


\section{Appendix I. Equation for the generalized energies}
In this appendix,  we show that generalized kinetic energy~($\KK$), generalized potential energy~($\PP$), generalized Hamiltonian~($\HH$), and generalized Lagrangian~($\LL$) satisfy differential-difference equation~\eq{Dyn kappa unif}.

We introduce matrix, $\ZZ$, consisting of covariances of
particle accelerations:
\be{zeta}
  \ZZ = \frac{1}{2}\MM^{\frac{1}{2}}\av{\ddot{\u}(\xx) \ddot{\u}(\yy)^{\top}}\MM^{\frac{1}{2}}.
\ee
Calculation of the second time derivatives of~$\KK$ and $\ZZ$ taking into
account equations of motion, yields:
\be{Dx Dy}
\begin{array}{l}
 \ddot{\KK} = \Dk_x \KK + \KK \Dk_y^{\top} +  2\ZZ,
 \qquad
\ddot{\ZZ} = \Dk_x \ZZ + \ZZ\Dk_y^{\top} + 2\Dk_x \KK \Dk_y^{\top},
\\[4mm]
 \DS \Dk_x \KK = \suma \MM^{-\frac{1}{2}}\CC_{\alpha}\MM^{-\frac{1}{2}} \KK\(\xx+ \va, \yy\),
 \quad
 \KK\Dk_y^{\rm T}  = \suma \KK\(\xx, \yy - \va\) \MM^{-\frac{1}{2}} \CC_{\alpha} \MM^{-\frac{1}{2}}.
 \end{array}
\ee
Excluding~$\ZZ$ from this  system of equations, we obtain
\be{Dyn kappa pril}
 \ddddot{\KK} -2\(\Dk_x\ddot{\KK}+\ddot{\KK}\Dk_y^{\top}\)
 + \Dk_x^2 \KK - 2\Dk_x\KK \Dk_y^{\top} + \KK \(\Dk_y^{\top}\)^2 = 0,
\ee
where $\Dk_x^2=\Dk_x\Dk_x$. Formula~\eq{Dyn kappa pril} exactly determines evolution of generalized kinetic energy for {\it any} initial conditions.

We consider initial conditions~\eq{IC matrix0}, corresponding to
spatially uniform temperature distribution. In this case the following identity is
satisfied~$\KK(\xx, \yy) = \KK\(\xx-\yy\)$. Using the identity we show that
\be{dxsapprox}
\begin{array}{l}
\DS \Dk_x \KK = \Dk \KK, \quad \KK\Dk_y^{\top} = \KK \Dk,
\quad
\Dk \KK \dfeq \suma \MM^{-\frac{1}{2}}\CC_{\alpha}\MM^{-\frac{1}{2}} \KK\(\xx-\yy + \va\).
\end{array}
\ee
Substitution of expressions~\eq{dxsapprox} for operators into equation~\eq{Dyn kappa pril}, yields equation~\eq{Dyn kappa unif}:
\be{Dyn kappa unif pril}
\DS
\ddddot{\KK} -2\(\Dk\ddot{\KK}+\ddot{\KK}\Dk\) + \Dk^2 \KK - 2\Dk\KK \Dk + \KK \Dk^2 = 0.
\ee

To derive equations for $\PI$, $\LL$, and $\HH$, we consider equation for~$\DD = \MM^{\frac{1}{2}}\av{\u(\xx) \u(\yy)^{\rm T}}\MM^{\frac{1}{2}}$. Calculation of the second time
derivatives of~$\DD$ and~$\KK$, yields
\be{K xi}
\ddot{\KK} = \Dk_x \KK + \KK \Dk_y^{\top} +  \Dk_x\DD\Dk_y^{\top},
\qquad
\ddot{\DD} = \Dk_x \DD + \DD \Dk_y^{\top} +  4\KK.
\ee
Excluding $\KK$ from this system, we obtain equation for~$\DD$:
\be{EQ xi}
 \ddddot{\DD} -2\(\Dk_x\ddot{\DD}+\ddot{\DD}\Dk_y^{\top}\)
 + \Dk_x^2 \DD - 2\Dk_x\DD \Dk_y^{\top} + \DD \(\Dk_y^{\top}\)^2 = 0.
\ee
Additionally, from formula~\eq{K xi} it also follows that
\be{}
 \LL = \frac{1}{4} \ddot{\DD}.
\ee
Thus~$\KK$ and $\DD$ satisfy the same linear equation~\eq{EQ xi}. Since generalized Hamiltonian, $\HH$, and generalized Lagrangian, $\LL$, are  linear functions of~$\KK$ and $\DD$, then they satisfy equation~\eq{EQ xi}. Trace of the generalized Hamiltonian is constant in time. Therefore~$\dev\HH$ also satisfies equation~\eq{EQ xi}.

\section{Appendix II. Dispersion relation}
In this appendix, the dispersion relation for a lattice, described by equations of motion~\eq{EM matr}, is derived.

We introduce new variable~$\UU(\xx) =\MM^{\frac{1}{2}} \u(\xx)$,
then equation of motion~\eq{EM matr} takes the form
\be{EM pr}
\begin{array}{l}
\DS
 \ddot{\UU}(\xx)  = \sum_{\alpha} \MM^{-\frac{1}{2}} \CC_{\alpha} \MM^{-\frac{1}{2}} \UU(\xx + \va),
\end{array}
\ee
where $\MM^{-\frac{1}{2}} \MM^{-\frac{1}{2}} = \MM^{-1}$.

The dispersion relation is derived by making substitution~$\UU = \Vect{A} e^{\ii\(\omega t + \kk\cdot\xx\)}$
in formula~\eq{EM pr}:
\be{Omega app}
 \(\Omb - \omega^2 \EE\)\Vect{A} = 0,
 \qquad
 \Omb =- \sum_{\alpha} \MM^{-\frac{1}{2}}\CC_{\alpha}\MM^{-\frac{1}{2}} e^{\ii\kk\cdot\va},
\ee
where $\EE$ is  identity matrix; $\Omb$ is referred to as the dynamical matrix~\cite{Dove}. Formula~\eq{Omega app} yields a homogeneous system of linear equations with respect to~$\Vect{A}$.
The system has nontrivial solution if the following condition is satisfied:
\be{det om}
 \det\(\Omb(\kk) - \omega^2 \EE\) = 0.
\ee
Here~$\det(..)$ stands for determinant of a matrix.
 Solutions of  equation~\eq{det om} are branches of dispersion relation~$\omega_j^2(\kk), j=1,..,N$. Note that from mathematical point of view,~$\omega_j^2$, are eigenvalues of~$\Omb$.

Finally, we show that dynamical matrix is Hermitian, i.e. it is equal to its own conjugate transpose:
\be{}
\begin{array}{l}
 \DS \Omb^{*\top}  = - \sum_{\alpha} \MM^{-\frac{1}{2}}\CC_{\alpha}^{\top}\MM^{-\frac{1}{2}} e^{-\ii\kk\cdot\va}
 \DS = -\sum_{\alpha} \MM^{-\frac{1}{2}}\CC_{-\alpha}\MM^{-\frac{1}{2}} e^{\ii\kk\cdot\Vect{a}_{-\alpha}} = \Omb,
 \end{array}
\ee
where identities~$\va = -\Vect{a}_{-\alpha}$, $\CC_{-\alpha} = \CC_{\alpha}^{\top}$ were used.

\section{Appendix III. Additional conservation laws for the generalized Hamiltonian}
In this appendix, we show that the generalized Hamiltonian satisfies additional conservation laws.

We introduce generalized potential energy,~$\PI(\xx,\yy)$, generalized Hamiltonian,~$\HH(\xx,\yy)$,
and generalized Lagrangian,~$\LL(\xx,\yy)$:
\be{HL comp pril}
\begin{array}{l}
  \DS \HH = \KK + \PI, \qquad \Tens{L} = \KK - \PI,
  \qquad
  \PI= -\frac{1}{4}\(\Dk_x \DD + \DD \Dk_y^{\rm T}\), \\[4mm]
\DS  \DD = \MM^{\frac{1}{2}}\av{\u(\xx) \u(\yy)^{\rm T}}\MM^{\frac{1}{2}}.
\end{array}
\ee
Here~$\KK=\KK(\xx,\yy)$; operators~$\Dk_x, \Dk_y$ are defined by formula~\eq{Dx Dy}.

Calculating time derivative of the generalized Hamiltonian taking into
account equations of motion, yields:
\be{dot H}
 \dot{\HH} = \frac{1}{4}\(\Dk_x \Vect{W}  - \Vect{W} \Dk_y^{\rm T}\),
 \qquad
 \Vect{W} = \MM^{\frac{1}{2}}\av{\u(\xx) \v(\yy)^{\rm T} - \vv(\xx)\u(\yy)^{\rm T}}\MM^{\frac{1}{2}}.
\ee
In the case of spatially uniform initial conditions~$\Vect{W} = \Vect{W}(\xx-\yy)$ and
$\Dk_x\Vect{W} = \Dk\Vect{W}$, $\Vect{W}\Dk_y^{\rm T}=\Vect{W}\Dk$. Then
\be{dH}
 \dot{\HH} = \frac{1}{4}\(\Dk\Vect{W} -\Vect{W}\Dk\).
\ee
Multiplying equation~\eq{dH} by $\Dk^n$, calculating trace and using identity~$\tr\(\Tens{A}\Tens{B}\)
= \tr\(\Tens{B}\Tens{A}\)$, yields conservation laws
\be{CLaws pril}
 \tr\(\Dk^n \HH\) = \tr\(\Dk^n \HH_0\),   \qquad   n=0,1,2,..,
\ee
where~$\HH_0$ is initial value of the generalized Hamiltonian.
For $n = 0$ and $\xx=\yy$, formula~\eq{CLaws pril} corresponds to conventional law of energy conservation. Similar conservation laws are derived for a one-dimensional chain in paper~\cite{Boldrighini1983} and for two- and three-dimensional monoatomic crystals in paper~\cite{Kuzkin_FTT}.

Formula~\eq{CLaws pril} can be written for trace and deviator of the generalized Hamiltonian:
\be{}
 \tr\HH = \tr\HH_0,   \qquad  \tr\(\Dk^n \dev\HH\) = \tr\(\Dk^n \dev\HH_0\),
 \qquad n=0,1,2...
\ee


\end{document}

%% file: def_Kuzkin.tex
\def\({\left(}
\def\){\right)}
\def\DS{\displaystyle}
\def\tr{{\rm tr}}
\def\av#1{\Bigl\langle{#1}\Bigr\rangle}
\def\be#1{\begin{equation}\label{#1}}
\def\ee{\end{equation}}
\def\BM{\begin{bmatrix}}
\def\EM{\end{bmatrix}}
\def\diag{{\rm diag}}

\def\eq#1{(\ref{#1})}

\def\xx{\Vect{x}}

\def\suma{\sum_{\alpha}}

\def\TT{\Vect{T}}
\def\DD{\Vect{D}}

\def\KKh{\hat{\KK}}

\def\kk{\Vect{k}}
\def\yy{\Vect{y}}
\def\vv{\Vect{v}}

\def\Omb{\mathbf{\Omega}}
\def\Lamb{\mathbf{\Lambda}}

\def\ve{\Vect{e}}
\def\va{\Vect{a}_{\alpha}}

\def\paptit#1{}

\def\Xi{\Tens{\xi}{}}

\def\MM{\Tens{M}{}}

\def\Gamma{\Tens{\gamma}{}}
\def\PP{\Tens{P}{}}

\def\ZZ{\Vect{Z}}

\def\ii{{\rm i}}

\def\dev{{\rm dev}}

\def\xx{\Vect{x}}

\def\dev{\rm dev}

\def\Dk{\Tens{\cal{L}}}

\def\EE{\Tens{E}}
\def\KK{\Tens{K}}

\def\u{\Vect u}

\def\diag{{\rm diag}}
\def\LL{\Tens{L}}
\def\CC{\Tens{C}}
\def\UU{\Vect{U}}
\def\BB{\Vect{B}}

\def\yy{\Vect{y}}
\def\vv{\Vect{v}}

\def\va{\Vect{a}_{\alpha}}

\newcommand{\Vect}[1]{\boldsymbol{\mathrm{#1}}{}}
\newcommand{\Tens}[1]{\boldsymbol{\mathrm{#1}} {}}
\def\be#1{\begin{equation}\label{#1}}
\def\ee{\end{equation}}
\def\eq#1{(\ref{#1})}

\def\oa{\end{array} $$}

\let\DS=\displaystyle

\def\({\left(}
\def\){\right)}

\let\DS     = \displaystyle

\def\v {\Vect{v}}

\def\vv{\Vect{v}}

\def\u{\Vect u}
\def\va{\Vect{a}_{\alpha}}

\def\paptit#1{}

\def\dev{{\rm dev}}

\def\xx{\Vect{x}}

\def\EE{\Tens{E}}

\def\yy{\Vect{y}}
\def\vv{\Vect{v}}

\def\ve{\Vect{e}}

\def\va{\Vect{a}_{\alpha}}

\def\[{\left[}
\def\]{\right]}
\def\({\left(}
\def\){\right)}
\def\av#1{\Bigl\langle{#1}\Bigr\rangle}

\def\av#1{\Bigl\langle{#1}\Bigr\rangle}
\def\dfeq{\,\stackrel{\mbox{\scriptsize def}}{=}\,}

\def\<{\langle}
\def\>{\rangle}

\def\Xi{\Tens{\xi}{}}

\def\KK{\Tens{K}{}}

\def\LL{\Tens{L}}
\def\HH{\Tens{H}}
\def\PI{\Tens{\Pi}} 